\documentclass[reqno, a4paper]{amsart}
\usepackage{amsmath}
\usepackage{amssymb}
\usepackage{amsthm}
\usepackage{xcolor}

\usepackage[scale=0.85]{geometry}

\usepackage{natbib}
\usepackage{bibentry} 

\usepackage[english]{babel}
\usepackage[utf8]{inputenc}

\usepackage{subfig}
\usepackage{graphicx}

\usepackage[unicode, breaklinks]{hyperref}
\usepackage[active]{srcltx} 

\usepackage{mathbbol}
\usepackage{bm} 
\usepackage{MnSymbol} 
\usepackage{gensymb}
\usepackage{eurosym}

\usepackage{units}
\usepackage{tensor}
\usepackage{accents}

\usepackage{enumitem}

\usepackage{lineno}

\usepackage{soul}

\DeclareMathOperator{\divergence}{div}

\DeclareMathOperator{\Tr}{Tr}

\newcommand{\exponential}[1]{\ensuremath{{\mathrm e}^{#1}}}

\newcommand{\reference}{\mathrm{ref}}

\newcommand{\bydefinition}{\mathrm{def}}
\newcommand{\traceless}[1]{{#1}_{\delta}}

\newcommand{\diff}{\mathrm{d}}

\renewcommand{\vec}[1]{\ensuremath{\mathbf{#1}}}

\makeatletter
\@ifpackageloaded{bm}%
{\renewcommand{\vec}[1]{\ensuremath{\bm{#1}}}%
}{%
\relax
}
\makeatother
\newcommand{\tensorq}[1]{\ensuremath{\mathbb{#1}}}      
\newcommand{\tensorc}[1]{\ensuremath{\mathrm{#1}}}      

\newcommand{\transpose}[1]{#1^\top}
\newcommand{\transposei}[1]{#1^{-\top}}
\newcommand{\inverse}[1]{#1^{-1}}

\newcommand{\identity}{\ensuremath{\tensorq{I}}}

\newcommand{\cstress}{\tensorq{T}}

\newcommand{\ecstress}{\tensorq{S}}

\newcommand{\fgrad}{\tensorq{F}}

\newcommand{\rcg}{\tensorq{C}}

\newcommand{\lcg}{\tensorq{B}}

\makeatletter
\@ifpackageloaded{bm}%
{%
}{%

}

\@ifpackageloaded{bm}%
{%
 
}{%

}

\makeatother

\newcommand{\generictensor}{{\tensorq{A}}}

\newcommand{\vecv}{\ensuremath{\vec{v}}}

\newcommand{\gradsym}{\ensuremath{\tensorq{D}}}

\newcommand{\gradvl}{\ensuremath{\tensorq{L}}}

\newcommand{\vhatp}[1][\vecvc]{{#1}^{\hat{\varphi}}}

\newcommand{\hatz}{\hat{z}}
\newcommand{\hatr}{\hat{r}}
\newcommand{\hatp}{\hat{\varphi}}

\newcommand{\cobvec}[1]{\vec{g}_{#1}} 


\newcommand{\ienergy}{\ensuremath{e}} 
\newcommand{\fenergy}{\ensuremath{\psi}} 
\newcommand{\entropy}{\ensuremath{\eta}} 

\newcommand{\temp}{\ensuremath{\theta}} 
\newcommand{\mns}{\ensuremath{m}} 

\newcommand{\cheatvol}{\ensuremath{c_{\mathrm{V}}}}

\newcommand{\efluxc}{\vec{j}_{e}} 

\newcommand{\hfluxc}{\vec{j}_{q}}     

\newcommand{\entfluxc}{\vec{j}_{\entropy}} 

\newcommand{\entprodc}{\xi} 
\newcommand{\entprodctemp}{\zeta} 

\newcommand{\pd}[2]{\ensuremath{\frac{\partial {#1}}{\partial {#2}}}}
\newcommand{\ppd}[2]{\ensuremath{\frac{\partial^2 {#1}}{\partial {#2^2}}}}
\newcommand{\dd}[2]{\ensuremath{\frac{\diff {#1}}{\diff {#2}}}}

\newcommand{\ddd}[2]{\ensuremath{\frac{\diff^2 {#1}}{\diff {#2}^2}}}

\newcommand{\fid}[1]{\ensuremath{\accentset{\triangledown}{#1}}}

\makeatletter
\@ifpackageloaded{tensor}
{

}{%

}
\makeatother

\makeatletter
\@ifpackageloaded{tensor}
{

}{%

}
\makeatother

\newcommand{\absnorm}[1]{\ensuremath{\left|#1\right|}}

\makeatletter
\@ifundefined{volume}{%
}%
{%
}
\makeatother

\makeatletter
\@ifpackageloaded{MnSymbol} 
{
\newcommand{\tensordot}[2]{\ensuremath{#1 \vdotdot #2}} 
}{%
\newcommand{\tensordot}[2]{\ensuremath{#1 : #2}} 
}
\makeatother
\newcommand{\vectordot}[2]{\ensuremath{#1 \bullet #2}}

\newcommand{\nplacer}{\kappa_{\mathnormal{p}(\mathnormal{t})}}  

\newcommand{\fgradrng}{\ensuremath{\tensorq{G}}} 

\newcommand{\gradvlrn}{\ensuremath{\gradvl_{\nplacer}}} 
\newcommand{\gradsymrn}{\ensuremath{\gradsym_{\nplacer}}} 

\newcommand{\lcgnc}{\ensuremath{\lcg_{\nplacer}}} 
\newcommand{\lcgncc}{\ensuremath{\lcgc_{\nplacer}}} 

\newcommand{\rcgnc}{\ensuremath{\rcg_{\nplacer}}} 

\newcommand{\fgradnc}{\ensuremath{\fgrad_{\nplacer}}} 
\newcommand{\fgradrc}{\ensuremath{\fgrad}} 

\newcommand{\sheatcapvol}{\cheatvol}

\renewcommand{\lcgncc}{\tensorc{B}}

\newcommand{\Brrhat}{\tensor{\lcgncc}{^\hatr_\hatr}}
\newcommand{\Brphat}{\tensor{\lcgncc}{^\hatr_\hatp}}

\newcommand{\Bprhat}{\tensor{\lcgncc}{^\hatp_\hatr}}
\newcommand{\Bpphat}{\tensor{\lcgncc}{^\hatp_\hatp}}

\newcommand{\Bzzhat}{\tensor{\lcgncc}{^\hatz_\hatz}}

\newcommand{\hencky}{\ensuremath{\varepsilon}}

\newcommand{\dhencky}{\ensuremath{\dot{\hencky}}}

\newcommand{\ddhencky}{\ensuremath{\ddot{\hencky}}}

\numberwithin{equation}{section}

\title[Viscoelastic fluids with temperature dependent material coefficients]{On thermodynamics of viscoelastic rate type fluids with temperature dependent material coefficients}

\author{Jaroslav Hron}
\address{
Faculty of Mathematics and Physics\\
Charles University\\
Sokolovsk\'a 83\\
Praha 8 -- Karl\'{\i}n\\
CZ 186\;75\\
Czech Republic
}
\email{hron@karlin.mff.cuni.cz}

\author{Vojt\v{e}ch Milo\v{s}}
\address{
Faculty of Mathematics and Physics\\
Charles University\\
Sokolovsk\'a 83\\
Praha 8 -- Karl\'{\i}n\\
CZ 186\;75\\
Czech Republic
}
\email{vojta.milos@gmail.com}

\author{V\'{\i}t Pr\r{u}\v{s}a}
\date{\today}
\address{
Faculty of Mathematics and Physics\\
Charles University\\
Sokolovsk\'a 83\\
Praha 8 -- Karl\'{\i}n\\
CZ 186\;75\\
Czech Republic
}
\email{prusv@karlin.mff.cuni.cz}

\author{Ond\v{r}ej Sou\v{c}ek}
\address{
Faculty of Mathematics and Physics\\
Charles University\\
Sokolovsk\'a 83\\
Praha 8 -- Karl\'{\i}n\\
CZ 186\;75\\
Czech Republic
}
\email{soucek@karel.troja.mff.cuni.cz}

\author{Karel T\r{u}ma}
\address{
Institute of Fundamental Technological Research\\
Polish Academy of Sciences\\
Adolfa Pawi\'nskiego 5B, Warszawa, PL 02-106, Poland
}
\email{ktuma@ippt.pan.pl}


\begin{document}

\begin{abstract}
We derive a class of thermodynamically consistent variants of Maxwell/Oldroyd-B type models for viscoelastic fluids. In particular, we study the models that allow one to consider temperature dependent material coefficients. This naturally calls for the formulation of a temperature evolution equation that would accompany the evolution equations for the mechanical quantities. The evolution equation for the temperature is explicitly formulated, and it is shown to be consistent with the laws of thermodynamics and the evolution equations for the mechanical quantities. The temperature evolution equation contains terms that are ignored or even not thought of in most of the works dealing with this class of fluids. The impact of the additional terms in the temperature evolution equation on the flow dynamics is documented by the solution of simple initial/boundary value problems.
\end{abstract}

\maketitle

\tableofcontents



\section{Introduction}
\label{sec:introduction-1}

Many viscoelastic fluid like materials exhibit strong temperature dependent behaviour. In most experimental studies, the temperature dependent behaviour is characterised in terms of the time-temperature superposition in the spirit of Williams--Landel--Ferry equation, see~\cite{williams.ml.landel.rf.ea:temperature}. (See also the early studies by~\cite{schwarzl.f.staverman.aj:time} and~\cite{wagner.mh.laun.hm:nonlinear} and other abundant contemporary studies.) The time-superposition factor provides one a characterisation of the temperature induced changes in creep and stress relaxation response, and its identification is undoubtedly an important step in the characterisation of the material. However, it is only a partial step in the formulation of the \emph{field equations} governing the motion of the given material. 

In particular, if the material under consideration is a viscoelastic material, then it is not completely clear what is the correct \emph{evolution equation for the temperature}. In fact, the question as ``how to account for the temperature changes'' is perceieved to be an open issue in rheology, see~\cite{tanner.ri:changing}.  Indeed, if the correct evolution equation for the temperature is absent, there is no chance to computationally model many important processes involving viscoelastic fluids. Clearly, this is a serious problem, and it deserves to be resolved. In what follows we present a possible \emph{solution to the problem in the case of incompressible Maxwell/Oldroyd-B type viscoelastic fluids with temperature dependent material coefficients}.

\section{Challenges in the formulation of the evolution equation for temperature}
\label{sec:remarks-temp-evol}
The formulation of the evolution equation for temperature $\temp$ is not as easy as it might seem. At the first glance the temperature evolution equation naturally arises via a straightforward reformulation of the generic evolution equation for the internal energy~$\ienergy$,
\begin{equation}
  \label{eq:18}
  \rho \dd{\ienergy}{t} = \tensordot{\cstress}{\gradsym} - \divergence \efluxc.
\end{equation}
(Here $\cstress$ denotes the Cauchy stress tensor, $\gradsym$ denotes the symmetric part of the velocity gradient, $\rho$ is the density and $\efluxc$ is the energy flux.) However, rewriting~\eqref{eq:18} in terms of temperature faces \emph{in the case of viscoelastic fluids} several challenges. 

First, unlike in the standard case of Navier--Stokes fluid, the specific internal energy~$\ienergy$ of a viscoelastic fluid consists of a \emph{thermal} and an \emph{elastic} contribution. This means that \emph{the internal energy is no longer a function of the temperature only}. In particular, formulae of the type $\ienergy = \sheatcapvol \temp$, where $\sheatcapvol$ denotes the specific heat capacity at constant volume, are no longer valid. This complicates the left hand side of the energy equation.

Second, the Cauchy stress tensor $\cstress$ in a viscoelastic fluid has a contribution generated by the viscous and elastic ``part'' of the fluid. However, the \emph{elastic part of the fluid does not}, by definition, \emph{dissipate the energy}. This implies that the product $\tensordot{\cstress}{\gradsym}$ does not provide a characterisation of the dissipation in the material. Indeed, the stress $\cstress$ necessarily contains a non-dissipative part that does not vanish in the product $\tensordot{\cstress}{\gradsym}$. In particular, the term $\tensordot{\cstress}{\gradsym}$ is not always positive as in the standard incompressible Navier--Stokes fluid where the positivity of the viscosity implies $\tensordot{\cstress}{\gradsym} = 2 \nu \tensordot{\gradsym}{\gradsym} \geq 0$. Consequently, in a viscoelastic fluid the internal energy can be partially converted to the kinetic energy. (Indeed, the \emph{elastic} part of the internal energy can be released and converted to the kinetic energy and the other way round.) Such a behaviour is not possible if one deals with the classical incompressible Navier--Stokes fluid where the internal energy has only a thermal part. This complicates the analysis of the right hand side of the energy equation. 

Third, the stresses generated by the elastic part should be consistent with the elastic contribution to the internal energy. For example, in the classical finite elasticity, the stress is related to a potential, namely to the free energy. (See for example~\cite{carroll.mm:must}.) Preferably, such a relation should also hold if one deals with the elastic part of the response of a viscoelastic fluid. This implies that \emph{the formula for the internal energy and the formula for the Cauchy stress tensor must be in some way consistent}. Consequently, an additional coupling between the mechanical and thermal quantities must appear.

Fourth, all the issues outlined above become more pronounced if the material coefficients characterising the mechanical response of the fluid are temperature dependent. If the description of the energy transfer mechanisms in the material is deficient, then the temperature evolution is necessarily predicted incorrectly. This can induce a significant error in the values of all temperature dependent material coefficients. Consequently, the mechanical part of the system of governing equations can be seriously corrupted as well, and the simulations based on such incomplete model can lead to false predictions. 

Finally, all the energy transfer mechanisms predicted by the chosen form of the internal energy and the Cauchy stress tensor must be consistent with the laws of thermodynamics. In particular, the chosen form of the internal energy and the Cauchy stress tensor must not lead to predictions that violate the second law of thermodynamics.

However, the frequently adopted approach to the temperature equation, especially amongst practitioners focused on numerical simulations, is far less subtle, see for example \cite{harder.h:numerical},  \cite{li.z.khayat.re:three-dimensional}, \cite{del-negro.c.currenti.g.ea:temperature-dependent}, \cite{choudhary.mk.venuturumilli.r.ea:mathematical}, \cite{salm.m.lucke.m:viscoelastic}, \cite{thielmann.m.kaus.bjp.ea:lithospheric} and \cite{cao.w.min.z.ea:numerical}. The temperature equation is usually written in the form 
\begin{equation}
  \label{eq:17}
  \rho \sheatcapvol \dd{\temp}{t} = \divergence(\kappa \nabla \temp),
\end{equation}
where $\kappa$ denotes the thermal conductivity. (Alternatively, the frictional heating term $2 \nu \tensordot{\gradsym}{\gradsym}$ can be also added to the right hand side of~\eqref{eq:17} as well.) This corresponds to a rather daring generalisation of the standard temperature evolution equation valid for the incompressible Navier--Stokes fluid. It is by no means granted that such a naive generalisation provides a reasonable description of the temperature evolution in a viscoelastic fluid. While such an approach \emph{might} provide a practically acceptable approximation of the true temperature evolution equation, it provides no insight into the previously discussed issues. Moreover, if one wants to either refine the evolution equation for the temperature, or if one wants to carefully justify the approximation being made, a full model must be developed. 

Despite of its practical importance, a little attention has been so far paid to the problem of correct formulation of the temperature evolution equation. The notable exceptions are the works by~\cite{wapperom.p.hulsen.ma:thermodynamics}, \cite{ireka.ie.chinyoka.t:non-isothermal,ireka.ie.chinyoka.t:analysis} and~\cite{guaily.a:temperature}. In these works, unlike in the general thermodynamic theories, see for example~\cite{leonov.ai:nonequilibrium} and \cite{rajagopal.kr.srinivasa.ar:thermodynamic}, the evolution equation for the temperature is formulated in an explicit form suitable for numerical simulations. However, these works have been mainly focused on \emph{compressible} materials with \emph{temperature independent} material coefficients. Full thermodynamically consistent models for~\emph{incompressible} viscoelastic materials with \emph{general temperature dependent} material coefficients are still absent in the literature and must be developed. 

In fact, the only thermodynamically based models we are aware of are that discussed by~\cite{rao.ij.rajagopal.kr:thermodynamic} and the follow-up works by~\cite{kannan.k.rao.ij.ea:thermomechanical}, \cite{kannan.k.rajagopal.kr:thermomechanical,kannan.k.rajagopal.kr:simulation} and \cite{kannan.k.rao.ij.ea:thermodynamic}, who have developed models for non-isothermal flows of incompressible viscoelastic fluids in their effort to model complex phenomena such as crystallisation in polymers.  However, these works are restricted to a class of materials where the entropic equation of state has a special additive structure that effectively leads to the internal energy that is a \emph{function of only the temperature}. Consequently, the temperature evolution equation has a simple structure and the shear modulus is allowed to be only a \emph{linear} function of temperature. Further, the governing equations for the mechanical quantities do not match the governing equations used in Maxwell/Oldroyd-B type models. These are, from the current perspective, substantial limitations and must be relaxed.

\section{Outline}
\label{sec:outline}
The ongoing investigation of the thermodynamics of incompressible viscoelastic fluids will be based on the approach adopted by \cite{rajagopal.kr.srinivasa.ar:thermodynamic} and \cite{malek.j.rajagopal.kr.ea:on}, see also Section~\ref{sec:outline-procedure}. However, these works are focused on the \emph{mechanical} response of viscoelastic fluids, hence a suitable generalisation of the approach to the \emph{non-isothermal} setting must be made, see Section~\ref{sec:modif-case-temp}. Using the thermodynamical considerations, see Section~\ref{sec:incompr-maxw-b}, we then derive a complete system of governing equations for incompressible \hbox{Maxwell/Oldroyd-B} type viscoelastic fluids. In particular, we explicitly formulate the temperature evolution equation for these materials. 

The system of governing equations, see Section~\ref{sec:incompr-maxw-b} for details, reads
\begin{subequations}
  \label{eq:maxwell-oldroyd-temperature-dependent}
  \begin{align}
    \label{eq:22}
    \divergence \vec{v} &= 0, \\
    \label{eq:19}
    \rho \dd{\vec{v}}{t}
    &=
    \divergence \cstress + \rho \vec{b}, \\
    \label{eq:23}
    \nu_1 \fid{\overline{\lcgnc}} 
    + 
    \mu \left(\lcgnc - \identity\right) 
    &= 
    \tensorq{0}
    ,
  \end{align}
  and
  \begin{multline}
    \label{eq:24}
    \left[
      \rho \sheatcapvol^{\mathrm{iNSE}} 
      -
      \left[
        \frac{\temp}{2}
        \ddd{\mu}{\temp}
        \left(
          \Tr \lcgnc
          -
          3
          -
          \ln \det \lcgnc
        \right)
      \right]
    \right]
    \dd{\temp}{t}
    =
    2 \nu \tensordot{\traceless{\gradsym}}{\traceless{\gradsym}}
    +
    \divergence \left(\kappa \nabla \temp \right)
    \\
    +
    \temp\dd{\mu}{\temp}
    \tensordot{
      \traceless{\left(\lcgnc\right)}
    }
    {\traceless{\gradsym}}
    +
    \frac{\mu}{2 \nu_1}
    \left(
      \mu - \temp \dd{\mu}{\temp}
    \right)
    \left(
      \Tr \lcgnc + \Tr \left( \inverse{\lcgnc} \right) - 6
    \right),
  \end{multline}
  where the Cauchy stress tensor $\cstress$ is given by the formulae
  \begin{align}
    \label{eq:25}
    \cstress 
    &= 
    \mns \identity + \traceless{\cstress}, 
    \\
    \label{eq:26}
    \traceless{\cstress} 
    &= 
    2 \nu \traceless{\gradsym} + \mu \traceless{\left( \lcgnc \right)}.
  \end{align}
\end{subequations}
In~\eqref{eq:maxwell-oldroyd-temperature-dependent} the symbol $\vec{v}$ denotes the velocity field, $\gradsym=_{\bydefinition} \frac{1}{2} \left( \nabla \vecv + \transpose{(\nabla \vec{v})} \right)$ denotes the symmetric part of the velocity gradient, $\mns$ is the mean normal stress (pressure), and $\traceless{\generictensor} =_{\bydefinition} \generictensor - \frac{1}{3} \left(\Tr \generictensor\right) \identity$ denotes the traceless part of the corresponding tensor. (Note that in virtue of the incompressibility condition~\eqref{eq:22} one gets the identity $\traceless{\gradsym} = \gradsym$.)  Symbol $\rho$ denotes the (constant) density of the material, and symbols $\nu_1$, $\nu$, $\mu$, $\sheatcapvol^{\mathrm{iNSE}}$ and $\kappa$ denote material parameters that can possibly depend on temperature. Symbol $\vec{b}$ stands for the specific body force and $\lcgnc$ denotes the ``extra stress'' tensor characterising the elastic part of the fluid response, see Section~\ref{sec:kinem-evolv-natur} for details. Finally, $\dd{}{t}=_{\bydefinition} \pd{}{t} + \vectordot{\vec{v}}{\nabla}$ denotes the material time derivative, and 
 \begin{equation}
   \label{eq:56}
   \fid{\generictensor}=_{\bydefinition} \dd{\generictensor}{t} - \gradvl \generictensor - \generictensor \transpose{\gradvl},
 \end{equation}
where $\gradvl=_{\bydefinition} \nabla \vec{v}$ denotes the upper convected derivative, see~\cite{oldroyd.jg:on}, and symbol $\tensordot{\tensorq{A}}{\tensorq{B}} = _{\bydefinition} \Tr \left(\tensorq{A} \transpose{\tensorq{B}}\right)$ denotes the scalar product on the space of matrices.

Upon the completion of system~\eqref{eq:maxwell-oldroyd-temperature-dependent} with appropriate boundary conditions, the system can be solved and used to determine the velocity~$\vec{v}$, pressure~$\mns$, temperature~$\temp$ and ``extra stress'' field~$\lcgnc$ in the given process of interest. As one might note, \emph{the governing equation~\eqref{eq:24} for the temperature is more complex than the simple heat equation~\eqref{eq:17}}.

The model introduced in~\eqref{eq:maxwell-oldroyd-temperature-dependent} is indeed a genuine generalisation of the standard Maxwell/Oldroyd-B model to the case of Maxwell/Oldroyd-B model with temperature independent material coefficients. The introduction of new unknown fields $p = _{\bydefinition} -m + \frac{\mu}{3} \Tr \lcgnc - \mu$, and $\ecstress =_{\bydefinition} \mu \left(\lcgnc - \identity\right)$ allows one to rewrite~\eqref{eq:23}, \eqref{eq:25} and~\eqref{eq:26} in an equivalent form
\begin{subequations}
  \label{eq:maxwell-oldroyd-temperature-dependent-solvent-polymer-form}
  \begin{align}
    \label{eq:28}
    \cstress &= -p \identity + 2\nu\gradsym + \ecstress, \\
    \label{eq:29}
    \nu_1 \fid{\overline{\left( \frac{\ecstress}{\mu} \right)}} + \ecstress &= 2 \nu_1 \gradsym.
  \end{align}
\end{subequations}
This is the formulation of the Maxwell/Oldroyd-B model frequently used in the mechanics of non-Newtonian fluids. (The total Cauchy stress~$\cstress$ is understood as a sum of the ``solvent'' contribution $2\nu\gradsym$ and the extra ``polymer'' contribution $\ecstress$.) Finally, yet another change of notation $\widetilde{\ecstress} =_{\bydefinition} \ecstress + 2 \nu \gradsym$ converts~\eqref{eq:maxwell-oldroyd-temperature-dependent-solvent-polymer-form} into the form
\begin{subequations}
  \label{eq:maxwell-oldroyd-temperature-dependent-original-oldroyd}
  \begin{align}
    \label{eq:30}
    \cstress &= -p \identity + \widetilde{\ecstress}, \\
    \label{eq:27}
    \nu_1 
    \fid{
      {
        \overline{ 
          \left(
            \frac{\widetilde{\ecstress}}{\mu}
          \right)
        }
      }
    }  
    + 
    \widetilde{\ecstress} 
    &= 
    2 \left(\nu_1 + \nu\right) \gradsym 
    + 
    2 \nu_1 
    \fid{
      \overline{
        \left( 
          \frac{\nu \gradsym}{\mu}
        \right)
      } 
    }.
  \end{align}
\end{subequations}
This is the formulation originally used by~\cite{oldroyd.jg:on}. 

Furthermore, in Section~\ref{sec:analyt-solut-govern}  we find semianalytical solutions to two simple boundary value problems for system~\eqref{eq:maxwell-oldroyd-temperature-dependent}. In Section~\ref{sec:cylindr-couette-flow-1} we solve the problem of the \emph{steady flow in between rotating coaxial cylinders with heated walls} (cylindrical Couette flow with heated walls). This problem is the classical flow problem extensively studied in rheology, see~\cite{couette.mma:etudes} and \cite{donnelly.rj:taylor-couette}. In Section~\ref{sec:time-depend-sque} we solve the problem of \emph{time dependent biaxial extension}, which is a setting relevant in the analysis of the squeeze flow, see~\cite{engmann.j.servais.c.ea:squeeze}. Again this is a classical flow problem extensively studied in rheology. The solutions of the problems are then used in the analysis of the impact of the various terms in the governing equations on the flow dynamics. 

Moreover, the semianalytical solutions also serve us in validating our implementation of a numerical solver for the system of governing equations, see Section~\ref{sec:numerical-solution}. Finally, using the numerical solver we simulate the behaviour of an incompressible viscoelastic material with temperature dependent material coefficients in a complex setting that leads to a non-uniform velocity, stress and temperature field. 

\section{Derivation of full thermodynamically consistent models}
\label{sec:derivation-model}
The goal is to derive a thermodynamically consistent variants of incompressible Maxwell/Oldroyd-B models with temperature dependent material coefficients. Here, the thermodynamical consistency means that the model must fulfill all the requirements discussed in Section~\ref{sec:remarks-temp-evol}. The derivation of the constitutive relations that is outlined below basically follows the procedure introduced by~\cite{rajagopal.kr.srinivasa.ar:thermodynamic}, \cite{rajagopal.kr.srinivasa.ar:on*7} and~\cite{malek.j.rajagopal.kr.ea:on}. (See also~\cite{malek.j.prusa.v:derivation} and references therein.) \emph{However, the referred works have been mainly focused on the isothermal setting, thus several important modifications of the procedure are at place}.

\subsection{General phenomenological approach}
\label{sec:outline-procedure}
The starting point of the phenomenological approach by~\cite{rajagopal.kr.srinivasa.ar:on*7} is the characterisation of the energy storage and entropy production mechanisms in the material. This is achieved by the identification of two scalar quantities, namely the specific internal energy $\ienergy$ and the specific entropy production $\entprodc$. The former quantity characterises the storage mechanisms, while the latter characterises the entropy production mechanisms. Naturally, the entropy production $\entprodc$ is chosen to be a non-negative function, which in return guarantees the consistence of the final set of constitutive relations with the second law of thermodynamics. 

The choice of these two \emph{scalar} quantities is then shown to imply the constitutive relations even for the \emph{tensorial} quantities such as the Cauchy stress tensor. The identification of the constitutive relation for the tensorial quantities from the knowledge of $\ienergy$ and $\entprodc$ goes as follows. The specification of the internal energy is provided by a formula of the type
\begin{equation}
  \label{eq:31}
  \ienergy = \ienergy(\entropy, y_1, \dots, y_n).
\end{equation}
Here $\entropy$ denotes the entropy and  $\left\{ y_i \right\}_{i=1}^n$ are some additional variables. Exploiting~\eqref{eq:31} in the evolution equation for the internal energy~\eqref{eq:18} then allows one to formulate the evolution equation for the entropy. Indeed, if one employs the standard definition of the temperature, 
\begin{equation}
  \label{eq:32}
  \temp =_{\bydefinition} \pd{\ienergy}{\entropy}(\entropy, y_1, \dots, y_n),
\end{equation}
then the application of the chain rule on the left hand side of~\eqref{eq:18} yields
\begin{equation}
  \label{eq:33}
  \temp \rho \dd{\entropy}{t} 
  +
  \rho \pd{\ienergy}{y_1} 
  \dd{y_1}{t}
  +
  \dots
  +
  \rho \pd{\ienergy}{y_n}
  \dd{y_n}{t}
  =
  \tensordot{\cstress}{\gradsym} - \divergence \hfluxc.
\end{equation}
(The energy flux $\efluxc$ in~\eqref{eq:18} has been identified with the heat flux $\hfluxc$.) The last equation is then rearranged into the form
\begin{equation}
  \label{eq:35}
  \rho \dd{\entropy}{t} + \divergence \left( \vec{g}(\entropy, \nabla \temp, \dots) \right)  = f(\entropy, \cstress, \gradsym, \hfluxc, \nabla \temp, \dots).
\end{equation}

Once this is done, the obtained equation is compared to the generic evolution equation for the entropy. The generic balance equation must take the form
\begin{equation}
  \label{eq:34}
  \rho \dd{\entropy}{t} + \divergence \left( \entfluxc \right) = \entprodc,
\end{equation}
where $\entfluxc$ denotes the entropy flux and $\entprodc$ denotes the entropy production. However, the entropy production $\entprodc$ is an already specified function, say $\entprodc = \entprodc (\gradsym, \nabla \temp, \dots)$. Consequently, the comparison of $f(\cstress, \gradsym, \hfluxc, \nabla \temp, \dots)$ on the right hand side of~\eqref{eq:35} with $\entprodc (\gradsym, \nabla \temp, \dots)$ on the right hand side of~\eqref{eq:34} yields the sought constitutive relations for the Cauchy stress tensor $\cstress$, the heat flux $\hfluxc$ and other quantities.

\subsection{Modifications of the general approach in the case of temperature dependent material coefficients}
\label{sec:modif-case-temp}
In the case of a material with temperature dependent coefficients, the procedure must be modified. Since we want to work with \emph{temperature} dependent material coefficients, it is no more convenient to work with the internal energy~$\ienergy$. The reason is that the natural variable of the internal energy is the entropy, see~\eqref{eq:31}, which naturally leads to entropy dependent material coefficients. However, it would be impractical to work with the \emph{entropy} dependent material coefficients. From the practical point of view, one needs \emph{temperature} dependent material coefficients, hence the internal energy must be replaced by another thermodynamical potential. 

In our case, the specific \emph{Helmholtz free energy} $\fenergy$ is a good choice. The specific Helmholtz free energy is a function of the temperature $\temp$ and other variables, see~\cite{callen.hb:thermodynamics}, and it is defined as a Legendre transform of internal energy with respect to the entropy
\begin{equation}
  \label{eq:36}
  \fenergy(\temp, y_1, \dots, y_n) 
  = 
  \ienergy(\entropy(\temp, y_1, \dots, y_n), y_1, \dots, y_n) 
  - 
  \temp \entropy(\temp, y_1, \dots, y_n),
\end{equation}
where $\entropy(\temp, y_1, \dots, y_n)$ is a function obtained by solving~\eqref{eq:32} for the entropy. The derivation of the constitutive relations is then, unlike in Section~\ref{sec:outline-procedure}, based on the identification of a formula for the free energy $\fenergy$ and the entropy production~$\entprodc$.

Having defined the free energy via~\eqref{eq:36} it follows that
\begin{equation}
  \label{eq:39}
  \entropy
  =
  -\pd{\fenergy}{\temp},
\end{equation}
hence the differentiation of~\eqref{eq:36} with respect to time yields
\begin{equation}
  \label{eq:37}
  \dd{\ienergy}{t}
  =  
  \temp
  \dd{\entropy}{t}
  +
  \pd{\fenergy}{y_1}
  \dd{y_1}{t}
  +
  \dots
  +
  \pd{\fenergy}{y_n}
  \dd{y_n}{t}
  .
\end{equation}
This is the sought expression for the time derivative of the internal energy $\ienergy$ in terms of the derivatives of the free energy $\fenergy$, the entropy $\entropy$ and other variables. Identity~\eqref{eq:37} can be used on the left hand side of the evolution equation for the internal energy~\eqref{eq:18}, which yields
\begin{equation}
  \label{eq:38}
  \rho \dd{\entropy}{t} + \divergence \left( \tilde{\vec{g}}(\temp, \nabla \temp, \dots) \right)  = \tilde{f}(\temp, \cstress, \gradsym, \hfluxc, \nabla \temp, \dots).
\end{equation}

This equation is the counterpart of~\eqref{eq:35}, and the rest of the procedure is identical to that outlined in Section~\ref{sec:outline-procedure}. The procedure allows one to identify the constitutive relations for the Cauchy stress tensor $\cstress$ and the heat flux $\hfluxc$. The benefit of using the free energy instead of the internal energy is that all the material coefficients in the constitutive relations are now directly expressed as functions of the temperature. However, an evolution equation for the temperature is still missing.

The sought evolution equation for the temperature is obtained by a simple manipulation. In~\eqref{eq:38} we have identified, with the help of the \emph{ansatz} for the entropy production $\entprodc$, the right hand side $\tilde{f}$ as well as the entropy flux $\tilde{\vec{g}}$. Consequently, if an explicit evolution equation for the temperature is needed, it remains to recall the relation between the free energy $\fenergy$ and the entropy $\entropy$, see~\eqref{eq:39}. Once~\eqref{eq:39} is substituted into~\eqref{eq:38}, we get
\begin{equation}
  \label{eq:40}
  -
  \rho \dd{}{t}\left( \pd{\fenergy}{\temp} \right) 
  + 
  \divergence \left( \tilde{\vec{g}} \right)  
  = 
  \entprodc
  .
\end{equation}

But the free energy $\fenergy$ and the entropy production $\entprodc$ are known quantities, they have been explicitly specified at the beginning of the procedure. Consequently, equation~\eqref{eq:40} and yet another application of the chain rule to the known function $\fenergy$ in~\eqref{eq:40} yield
\begin{equation}
  \label{eq:41}
  \frac{\rho \sheatcapvol}{\temp} \dd{\temp}{t}
  -
  \rho
  \pd{^2\fenergy}{y_1 \partial \temp}
  \dd{y_1}{t}
  +
  \dots
  -
  \rho
  \pd{^2\fenergy}{y_n \partial \temp}
  \dd{y_n}{t}
  + 
  \divergence \left( \tilde{\vec{g}} \right)  
  =
  \entprodc
  .
\end{equation}
where
\begin{equation}
  \label{eq:42}
  \sheatcapvol
  =_{\bydefinition}
  - \temp  \ppd{\fenergy}{\temp}
\end{equation}
is the heat capacity at constant volume, see~\cite{callen.hb:thermodynamics}. Equation~\eqref{eq:41} is the sought explicit evolution equation for the temperature.

\subsubsection{Comment on kinetic theory based approaches to the constitutive relations}
\label{sec:comm-kinet-theory}
The macroscopic models for the description of viscoelastic fluids are often derived by appealing to the kinetic theory type arguments, see for example~\cite{bhave.av.armstrong.rc.ea:kinetic}. Such a derivation of the macroscopic model is typically based on various approximations. For example, the inertia of the particles forming the polymer chains is neglected, drag on the particles is calculated via the Stokes formula, temperature field is assumed to be homogeneous and so forth. Yet the resulting macroscopic model arising from these approximations might still provide a useful insight into the interplay between microscopic and macroscopic characterisation of the given fluid. 

On the other hand, such a model can not be expected to provide \emph{consistent} characterisation of the energy transfer mechanisms in the sense of compatibility of the macroscopic governing equations for the mechanical and thermal quantities. (See Section~\ref{sec:remarks-temp-evol} for the discussion of the compatibility requirements.) The reason is that the mutual consistency of the approximations being made is simply too difficult or even impossible to achieve. The mutual inconsistency of the approximations can then result in inaccuracies in the description of energy transfer mechanisms, which is especially true if one would like to have a macroscopic model that works for \emph{strongly inhomogeneous temperature and velocity fields}. Consequently, the kinetic theory type approach hardly provides an easy to use alternative to the phenomenological approach outlined above. If the description of microscopic behaviour of the material is of no interest---which is the case in the investigation of the macroscopic \emph{flows}---then a purely phenomenological approach should be preferred.  

\subsection{Incompressible Maxwell/Oldroyd-B type models with temperature dependent material coefficients}
\label{sec:incompr-maxw-b}
Let us now use the general procedure in a specific case of a viscoelastic fluid. As argued by~\cite{rajagopal.kr.srinivasa.ar:thermodynamic} and~\cite{malek.j.rajagopal.kr.ea:on}, the motion of a viscoelastic fluid can be on the phenomenological level understood as follows. 

\subsubsection{Kinematics of evolving natural configuration}
\label{sec:kinem-evolv-natur}
\begin{figure}[h]
  \centering
  \includegraphics[width=0.4\textwidth]{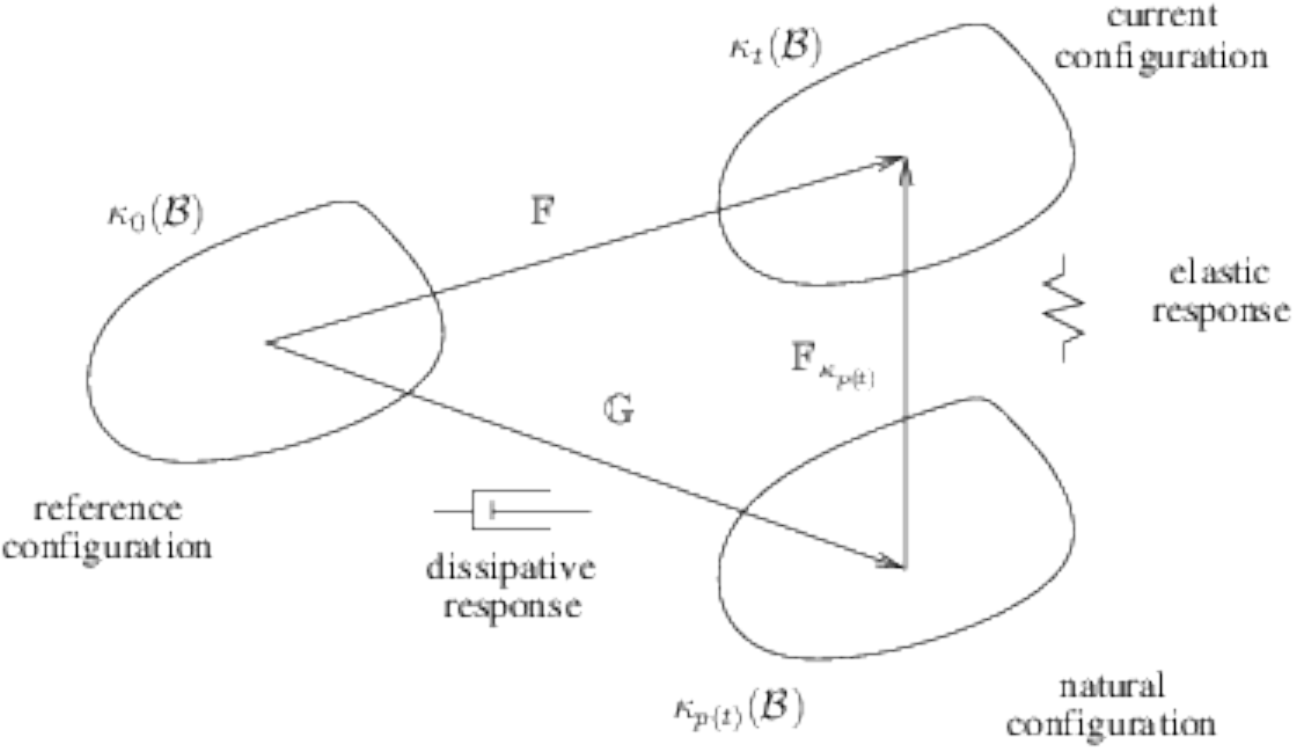}
  \caption{Viscoelastic fluid -- kinematics.}
  \label{fig:viscoelastic-kinematics}
\end{figure}

The deformation from the initial configuration to the current configuration is \emph{virtually} split to the deformation of the natural configuration and to the instantaneous elastic deformation from the natural configuration to the current configuration. The evolution of the natural configuration is understood as an entropy producing process. On the other hand, the energy storage ability is attributed to the elastic deformation from the natural configuration to the current configuration. (Figure~\ref{fig:viscoelastic-kinematics} depicts the situation, see also~\cite{prusa.rajagopal.kr:on*1} and \cite{malek.j.prusa.v:derivation} for details.) Such a decomposition is loosely motivated by the spring-dashpot model for the behaviour of Maxwell type viscoelastic fluid, see for example~\cite{wineman.as.rajagopal.kr:mechanical}.

If the total deformation is seen as a composition of the two deformations, then the total deformation gradient $\fgrad$ can be written~as
\begin{equation}
  \label{eq:43}
  \fgradrc = \fgradnc \fgradrng,
\end{equation}
where $\fgradnc$ and $\fgradrng$ are the deformation gradients of the partial deformations. Motivated by the standard relation between the spatial velocity gradient $\gradvl=_{\bydefinition} \nabla \vec{v}$ and the deformation gradient $\fgrad$, 
\begin{equation}
  \label{eq:44}
  \dd{\fgrad}{t} = \gradvl \fgrad,
\end{equation}
one introduces new tensorial quantities $\gradvlrn$ and $\gradsymrn$ defined as
\begin{equation}
  \label{eq:219}
  \gradvlrn =_{\bydefinition} \dd{\fgradrng}{t} \inverse{\fgradrng}
  ,
  \qquad
  \gradsymrn = _{\bydefinition} \frac{1}{2} \left(\gradvlrn + \transpose{\gradvlrn}\right).
\end{equation}
Using $\gradvlrn$, one can express the material time derivative of $\fgradnc$ as
\begin{equation}
  \label{eq:45}
  \dd{\fgradnc}{t}
  =
  \gradvl \fgradnc - \fgradnc \gradvlrn
  .
\end{equation}
Further, the material time derivative of the left Cauchy--Green tensor $\lcgnc =_{\bydefinition} \fgradnc \transpose{\fgradnc}$ associated to the instantaneous elastic (non-dissipative) response then reads
\begin{equation}
  \label{eq:46}
  \dd{\lcgnc}{t}
  =
  \gradvl \lcgnc
  +
  \lcgnc \transpose{\gradvl}
  -
  2 \fgradnc \gradsymrn \transpose{\fgradnc}.
  .
\end{equation}
Note that the last formula reduces, using the definition of the upper convected derivative \eqref{eq:56}, to the formula
\begin{equation}
  \label{eq:48}
  \fid{\overline{\lcgnc}} 
  =   
  -
  2 \fgradnc \gradsymrn \transpose{\fgradnc}.
\end{equation}

Since $\lcgnc$ represents the elastic (non-dissipative) part of the response, it should naturally enter the formula for the internal/free energy of the material. The reason is that the energy storage ability is in finite elasticity theory described in terms of the left Cauchy--Green tensor $\lcg =_{\bydefinition} \fgrad \transpose{\fgrad}$. But in our case only a part of the total deformation is non-dissipative/elastic. Consequently, only $\lcgnc =_{\bydefinition} \fgradnc \transpose{\fgradnc}$ instead of $\lcg$ then plays the role of an additional variable in the formula for the internal energy~\eqref{eq:31} or the free energy~\eqref{eq:36}. 

\subsubsection{Ansatz for the free energy $\fenergy$ and the evolution equation for the entropy}
\label{sec:ansatz-free-energy}

Now we are in a position to choose the \emph{ansatz} for the free energy $\fenergy$, and proceed with the procedure outlined in Section~\ref{sec:modif-case-temp}. We are considering a homogeneous \emph{incompressible} viscoelastic fluid that does not change volume even if it is a subject to thermal stimuli. This means that the density $\rho$ does not enter the formula formula for the energy,
\begin{equation}
  \label{eq:49}
  \fenergy(\temp, \rho, \lcgnc) = \overline{\fenergy}(\temp, \lcgnc).
\end{equation}
In particular, the \emph{ansatz} for the free energy takes the form
\begin{equation}
  \label{eq:50}
  \overline{\fenergy}(\temp, \lcgnc)
  =_{\bydefinition}
  \widetilde{\fenergy} \left(\temp\right)
  + 
  \frac{\mu(\temp)}{2\rho}
  \left(
    \Tr \lcgnc
    -
    3
    -
    \ln \det \lcgnc
  \right)
\end{equation}
where $\mu(\temp)$ is a positive function of the temperature, and $\widetilde{\fenergy}$ is a positive function of the temperature. This is, up to the temperature dependence of $\mu$ and $\widetilde{\fenergy}$, the same \emph{ansatz} as in~\cite{malek.j.rajagopal.kr.ea:on}. The term 
$ 
\frac{\mu(\temp)}{2\rho}
\left(
  \Tr \lcgnc
  -
  3
  -
  \ln \det \lcgnc
\right)$  corresponds to the assumption that the elastic response from the natural to the current configuration is the response of a compressible neo-Hookean elastic material. The \emph{ansatz} has been shown, for constant $\mu$ and $\widetilde{\fenergy}$, to generate the standard Maxwell/Oldroyd-B model. (Provided that the entropy production $\entprodc$ is also chosen in an appropriate manner.) Consequently, the \emph{ansatz} \eqref{eq:50} is expected to lead to a variant of Maxewell/Oldroyd-B model with temperature dependent material coefficients. In particular, the model should be capable of describing temperature dependent elastic response.

Now we explicitly evaluate~\eqref{eq:37} for the \emph{ansatz}~\eqref{eq:50}. This yields
\begin{equation}
  \label{eq:54}
  \dd{\ienergy}{t}
  =
  \temp
  \dd{\entropy}{t}
  +
  \pd{\overline{\fenergy}}{\Tr \lcgnc}
  \dd{}{t}\left( \Tr \lcgnc \right)
  +
  \pd{\overline{\fenergy}}{\ln \left[ \det \lcgnc \right]}
  \dd{}{t}
  \left(
    \ln \left[ \det \lcgnc \right]
  \right)
  .
\end{equation}

Using~\eqref{eq:46} we find that the time derivatives of the trace and the determinant of the \emph{left} Cauchy--Green tensor $\lcgnc$ read
\begin{subequations}
  \label{eq:51}
  \begin{align}
    \label{eq:52}
    \dd{}{t} \Tr \lcgnc
    &=
    2 \tensordot{\lcgnc}{\gradsym}
    -
    2 \tensordot{\rcgnc}{\gradsymrn}
    ,
    \\
    \label{eq:53}
    \dd{}{t}
    \left(
      \ln \left[ \det \lcgnc \right]
    \right)
    &=
    2 \tensordot{\identity}{\gradsym}
    -
    2 \tensordot{\identity}{\gradsymrn}
    ,
  \end{align}
\end{subequations}
where $\rcgnc=_{\bydefinition} \transpose{\fgradnc} \fgradnc$ denotes the \emph{right} Cauchy--Green tensor associated to the elastic part of the deformation. Substitution of~\eqref{eq:51} into~\eqref{eq:54} then yields 
\begin{equation}
  \label{eq:55}
  \dd{\ienergy}{t}
  =
  \temp
  \dd{\entropy}{t}
  +
  \frac{\mu(\temp)}{\rho}
  \left(
    \tensordot{\lcgnc}{\gradsym}
    -
   \tensordot{\rcgnc}{\gradsymrn}
  \right)
  -
  \frac{\mu(\temp)}{\rho}
  \left(
    \tensordot{\identity}{\gradsym}
    -
    \tensordot{\identity}{\gradsymrn}
  \right)
  .
\end{equation}
Using this formula on the left hand side of the evolution equation for the internal energy~\eqref{eq:18} then gives the sought evolution equation for the entropy
\begin{equation}
  \label{eq:57}
  \temp
  \rho
  \dd{\entropy}{t}
  =
  \tensordot{\left( \traceless{\cstress} - \mu(\temp) \traceless{\left(\lcgnc\right)} \right)}{\traceless{\gradsym}}
  +
  \mu(\temp)
  \tensordot{\left(\rcgnc - \identity\right)}{\gradsymrn}
  -
  \divergence \hfluxc
  .
\end{equation}
Here we have used the fact that $\tensordot{\identity}{\gradsym} = \Tr \gradsym =0$, which implies that $\gradsym = \traceless{\gradsym}$. Further, we have also decomposed the Cauchy stress to its spherical and traceless part, $\cstress =_{\bydefinition} \mns \identity + \traceless{\cstress}$. 

The standard manipulation then leads to the equation that has the form~\eqref{eq:35}, namely
\begin{equation}
  \label{eq:58}
  \rho
  \dd{\entropy}{t}
  +
  \divergence 
  \left(
    \frac{\hfluxc}{\temp}
  \right)
  =
  \frac{1}{\temp}
  \left[
    \tensordot{\left( \traceless{\cstress} - \mu(\temp) \traceless{\left(\lcgnc\right)} \right)}{\traceless{\gradsym}}
    +
    \mu(\temp)
    \tensordot{\left(\rcgnc - \identity\right)}{\gradsymrn}
  \right]
  -
  \frac{\vectordot{\hfluxc}{\nabla \temp}}{\temp^2}.
\end{equation}
Note that the spherical part of the stress, that is $\mns \identity$, does not enter the right hand side of~\eqref{eq:58}. This is natural. The spherical part of the stress includes the force that prevents the fluid from changing its volume, and this ``constraint'' force does not produce the entropy.

\subsubsection{Ansatz for the entropy production $\entprodc$}
\label{sec:ansatz-entr-prod}
Let us now assume that the entropy production $\entprodc =_{\bydefinition} \frac{\entprodctemp}{\temp}$ is given by the formula 
\begin{equation}
  \label{eq:59}
  \entprodctemp
  =_{\bydefinition}
  2 \nu(\temp) \tensordot{\traceless{\gradsym}}{\traceless{\gradsym}}
  +
  2 \nu_1(\temp)
  \Tr 
  \left(
    \fgradnc \gradsymrn \transpose{\left( \fgradnc \gradsymrn \right)}
  \right)
  +
  \kappa(\temp)
  \frac{\absnorm{\nabla \temp}^2}{\temp}
  ,
\end{equation}
where the functions $\nu$, $\nu_1$ and $\kappa$ are positive functions of the temperature. (If necessary these can be also positive functions of $\temp$, $\lcgnc$ and $\gradsym$.) The chosen form of the entropy production guarantees that the entropy production is non-negative, and in the isothermal case it is known to lead to the standard Maxwell/Oldroyd-B model, see~\cite{malek.j.rajagopal.kr.ea:on}.

Comparison of the entropy production $\entprodc$ given by~\eqref{eq:59} with the right hand side of~\eqref{eq:58} yields the sought constitutive relations
\begin{subequations}
  \label{eq:constitutive-relations}
  \begin{align}
    \label{eq:61}
    \traceless{\cstress} - \mu(\temp) \traceless{\left(\lcgnc\right)} &= 2 \nu(\temp) \traceless{\gradsym}, \\
    \label{eq:47}
    \mu(\temp)
    \left(\rcgnc - \identity\right)
    &=
    \nu_1(\temp)  \left(\rcgnc \gradsymrn + \gradsymrn \rcgnc\right)
    ,
    \\
    \label{eq:60}
    \hfluxc &= - \kappa(\temp) \nabla \temp.
  \end{align}
\end{subequations}

\subsubsection{Evolution equation for the left Cauchy--Green tensor $\lcgnc$}
\label{sec:evol-equat-left}
Equation~\eqref{eq:47} must be further manipulated in order to get an evolution equation for the unknown tensor field~$\lcgnc$. Recall that the kinematical considerations, see~\eqref{eq:48}, led to the equation
\begin{equation}
  \label{eq:62}
  \fid{\overline{\lcgnc}} 
  =   
  -
  2 \fgradnc \gradsymrn \transpose{\fgradnc}.
\end{equation}
This equation can serve as an evolution equation for the left Cauchy--Green tensor $\lcgnc$ provided that one is able to rewrite the right hand side in terms of $\gradsym$ and $\lcgnc$. This is the point where~\eqref{eq:47} comes into play. First, if~\eqref{eq:47} holds, then~$\gradsymrn$ necessarily commutes with $\rcgnc$, see~\cite{rajagopal.kr.srinivasa.ar:thermodynamic} for a proof. Second, the multiplication of~\eqref{eq:47} from the left by $\transposei{\fgradnc}$ and by $\transpose{\fgradnc}$ from the right yields
\begin{equation}
  \label{eq:63}
   \mu(\temp)
   \left(\lcgnc - \identity\right)
   =
   2\nu_1(\temp) \fgradnc \gradsymrn \transpose{\fgradnc},
\end{equation}
which is the sought formula for the product $\fgradnc \gradsymrn \transpose{\fgradnc}$. Substituting~\eqref{eq:63} into~\eqref{eq:62} then leads to
\begin{equation}
  \label{eq:64}
  \nu_1(\temp)\fid{\overline{\lcgnc}} 
  +
  \mu(\temp)
  \left(\lcgnc - \identity\right)
  =
  \tensorq{0}.
\end{equation}
 
\subsubsection{Entropy production $\entprodc$ in terms of the left Cauchy--Green tensor $\lcgnc$}
\label{sec:entr-prod-terms}

The entropy production $\entprodc =_{\bydefinition} \frac{\entprodctemp}{\temp}$ \emph{ansatz} has been specified in terms of $\fgradnc$ and $\gradsymrn$ which is inconvenient. (We do not have an explicit evolution equation for any of these quantities.) As in the previous case, it is desirable to rewrite the entropy production in terms of the unknown fields~$\vec{v}$ and $\lcgnc$.  To achieve this, we need to find an alternative expression for the term $
\Tr 
\left(
  \fgradnc \gradsymrn \transpose{\left( \fgradnc \gradsymrn \right)}
\right)
$. This can be done as follows
\begin{equation}
  \label{eq:65}
  \Tr 
  \left(
    \fgradnc \gradsymrn \transpose{\left( \fgradnc \gradsymrn \right)}
  \right)
  =
  \Tr 
  \left(
    \fgradnc \gradsymrn \transpose{\fgradnc} \transposei{\fgradnc} \inverse{\fgradnc} \fgradnc \gradsymrn \transpose{\fgradnc}
  \right)
  =
  \frac{1}{4}
  \Tr 
  \left(
    \fid{\overline{\lcgnc}}
    \inverse{\lcgnc}
    \fid{\overline{\lcgnc}}
  \right),  
\end{equation}
where we have exploited the kinematical identity~\eqref{eq:62} and the definition of $\lcgnc$. Further, if we use~\eqref{eq:64} in~\eqref{eq:65}, we see that the term in the entropy production formula can be in fact rewritten as
\begin{equation}
  \label{eq:66}
  \Tr 
  \left(
    \fgradnc \gradsymrn \transpose{\left( \fgradnc \gradsymrn \right)}
  \right)
  =
  \frac{\mu^2(\temp)}{4 \nu_1^2(\temp)}
  \left(
    \Tr \lcgnc + \Tr \left( \inverse{\lcgnc} \right) - 6
  \right)
  .
\end{equation}

Consequently, the entropy production $\entprodc =_{\bydefinition} \frac{\entprodctemp}{\temp}$ can be equivalently written either as
\begin{equation}
  \label{eq:67}
  \entprodctemp
  =
  2 \nu(\temp) \tensordot{\traceless{\gradsym}}{\traceless{\gradsym}}
  +
  \frac{\nu_1(\temp)}{2}
  \Tr 
  \left(
    \fid{\overline{\lcgnc}}
    \inverse{\lcgnc}
    \fid{\overline{\lcgnc}}
  \right)
  +
  \kappa(\temp)
  \frac{\absnorm{\nabla \temp}^2}{\temp}
\end{equation}
or as
\begin{equation}
  \label{eq:68}
  \entprodctemp
  =
  2 \nu(\temp) \tensordot{\traceless{\gradsym}}{\traceless{\gradsym}}
  +
  \frac{\mu^2(\temp)}{2 \nu_1(\temp)}
  \left(
    \Tr \lcgnc + \Tr \left( \inverse{\lcgnc} \right) - 6
  \right)
  +
  \kappa(\temp)
  \frac{\absnorm{\nabla \temp}^2}{\temp}  
  .
\end{equation}
The first form immediately indicates that the entropy production depends on the rate quantity $\fid{\overline{\lcgnc}}$, while the second form is less complex and might be advantageous for numerical computations.
 
\subsubsection{Evolution equation for the temperature}
\label{sec:evol-equat-temp}
It remains to derive the evolution equation for the temperature. Following the procedure outlined in Section~\ref{sec:modif-case-temp}, we need to find the derivative of the free energy with respect to the temperature. Using the \emph{ansatz} \eqref{eq:50}, we see that
\begin{equation}
  \label{eq:69}
  \pd{\fenergy}{\temp}
  =
  \pd{\widetilde{\fenergy}}{\temp}
  +
  \frac{1}{2\rho}
  \dd{\mu}{\temp}
  \left(
    \Tr \lcgnc
    -
    3
    -
    \ln \det \lcgnc
  \right)
  .
\end{equation}
Substituting~\eqref{eq:69} into the evolution equation for the entropy~\eqref{eq:40} that for the given entropy flux $\entfluxc = \frac{\hfluxc}{\temp}$ and entropy production $\entprodc = \frac{\entprodctemp}{\temp}$ reads
\begin{equation}
  \label{eq:70}
  -
  \rho \dd{}{t}\left( \pd{\fenergy}{\temp} \right) 
  + 
  \divergence \left( \frac{- \kappa(\temp) \nabla \temp}{\temp} \right) 
  =
  \frac{1}{\temp}
  \left[
    2 \nu(\temp) \tensordot{\traceless{\gradsym}}{\traceless{\gradsym}}
    +
    \frac{\nu_1(\temp) \mu^2(\temp)}{2 \nu_1^2(\temp)}
    \left(
      \Tr \lcgnc + \Tr \left( \inverse{\lcgnc} \right) - 6
    \right)
    +
    \kappa(\temp)
    \frac{\absnorm{\nabla \temp}^2}{\temp} 
  \right]
  ,
\end{equation}
yields
\begin{multline}
\label{eq:75}
    \left[
      \rho \sheatcapvol^{\mathrm{iNSE}} 
      -
      \left[
        \frac{\temp}{2}
        \ddd{\mu}{\temp}
        \left(
          \Tr \lcgnc
          -
          3
          -
          \ln \det \lcgnc
        \right)
      \right]
    \right]
    \dd{\temp}{t}
    =
    2 \nu \tensordot{\traceless{\gradsym}}{\traceless{\gradsym}}
    +
    \divergence \left(\kappa \nabla \temp \right)
    \\
    +
    \temp\dd{\mu}{\temp}
    \tensordot{
      \traceless{\left(\lcgnc\right)}
    }
    {\traceless{\gradsym}}
    +
    \frac{\mu}{2 \nu_1}
    \left(
      \mu - \temp \dd{\mu}{\temp}
    \right)
    \left(
      \Tr \lcgnc + \Tr \left( \inverse{\lcgnc} \right) - 6
    \right).
  \end{multline}
In~\eqref{eq:75} we have introduced the notation
\begin{equation}
  \label{eq:76}
  \sheatcapvol^{\mathrm{iNSE}} =_{\bydefinition} - \temp \ppd{\widetilde{\fenergy}}{\temp}.
\end{equation} 
Further, we have also used the formulae for the time derivative of the trace and the determinant of the left Cauchy--Green tensor, see~\eqref{eq:51}, and the equalities
\begin{subequations}
  \label{eq:72}
  \begin{align}
    \tensordot{\rcgnc}{\gradsymrn} 
    &=
    \Tr 
    \left(
      \fgradnc \gradsymrn \transpose{\fgradnc}
    \right)
    =
    -\frac{1}{2}
    \Tr \left(  \fid{\overline{\lcgnc}} \right)
    =
    \frac{\mu(\temp)}{2 \nu_1(\temp)}
    \left(
      \Tr \lcgnc
      -
      3
    \right)
  ,
  \\
  \Tr \gradsymrn
  &=
  \Tr
  \left(
    \fgradnc
    \gradsymrn
    \transpose{\fgradnc}
    \transposei{\fgradnc}
    \inverse{\fgradnc}
  \right)
  =
  -\frac{1}{2}
  \Tr
  \left(
    \fid{\overline{\lcgnc}}
    \inverse{\lcgnc}
  \right)
  =
  \frac{\mu(\temp)}{2 \nu_1(\temp)}
  \left(
    3 - \Tr \left( \inverse{\lcgnc} \right)
  \right)
  .
  \end{align}
\end{subequations}
Equation~\eqref{eq:75} is the sought explicit evolution equation for the temperature.

\subsubsection{Summary}
\label{sec:summary}
Let us consider a homogeneous incompressible viscoelastic material characterised by the free energy $\fenergy$,
\begin{equation}
  \label{eq:free-energy-summary}
  \fenergy 
  =_{\bydefinition} 
  \widetilde{\fenergy} \left(\temp\right)
  + 
  \frac{\mu(\temp)}{2\rho}
  \left(
    \Tr \lcgnc
    -
    3
    -
    \ln \det \lcgnc
  \right)
\end{equation}
and the entropy production $\entprodc =\frac{\entprodctemp}{\temp}$, where
\begin{equation}
  \label{eq:entropy-production-summary}
  \entprodctemp
  =_{\bydefinition}
  2 \nu(\temp) \tensordot{\traceless{\gradsym}}{\traceless{\gradsym}}
  +
  \frac{\mu^2(\temp)}{2 \nu_1(\temp)}
  \left(
    \Tr \lcgnc + \Tr \left( \inverse{\lcgnc} \right) - 6
  \right)
  +
  \kappa(\temp)
  \frac{\absnorm{\nabla \temp}^2}{\temp}  
  .
\end{equation}
The chosen form of the free energy and the entropy production implies that the field equations governing the motion of the material read
\begin{subequations}
  \label{eq:maxwell-oldroyd-temperature-dependent-summary}
  \begin{align}
    \label{eq:74}
    \divergence \vec{v} &= 0, \\
    \label{eq:71}
    \rho \dd{\vec{v}}{t}
    &=
    \divergence \cstress + \rho \vec{b}, \\
    \label{eq:73}
    \nu_1(\temp) \fid{\overline{\lcgnc}} 
    + 
    \mu(\temp) \left(\lcgnc - \identity\right) 
    &= 
    \tensorq{0}
    ,
  \end{align}
  and
  \begin{multline}
    \label{eq:77}
    \left[
      \rho \sheatcapvol^{\mathrm{iNSE}}
      -
      \left[
        \frac{\temp}{2}
        \ddd{\mu}{\temp}(\temp)
        \left(
          \Tr \lcgnc
          -
          3
          -
          \ln \det \lcgnc
        \right)
      \right]
    \right]
    \dd{\temp}{t}
    =
    2 \nu(\temp) \tensordot{\traceless{\gradsym}}{\traceless{\gradsym}}
    +
    \divergence \left(\kappa(\temp) \nabla \temp \right)
    \\
    +
    \temp\dd{\mu}{\temp}
    \tensordot{
      \traceless{\left(\lcgnc\right)}
    }
    {\traceless{\gradsym}}
    +
    \frac{\mu(\temp)}{2 \nu_1(\temp)}
    \left(
      \mu(\temp) - \temp \dd{\mu}{\temp}(\temp)
    \right)
    \left(
      \Tr \lcgnc + \Tr \left( \inverse{\lcgnc} \right) - 6
    \right),
  \end{multline}
  where the Cauchy stress tensor $\cstress$ is given by the formulae
  \begin{equation}
    \label{eq:78}
    \cstress 
    = 
    \mns \identity + \traceless{\cstress}, 
    \qquad
    \traceless{\cstress} 
    = 
    2 \nu(\temp) \traceless{\gradsym} + \mu(\temp) \traceless{\left( \lcgnc \right)}.
  \end{equation}
\end{subequations}
System~\eqref{eq:maxwell-oldroyd-temperature-dependent-summary} provides a closed system of equations for the unknown velocity field $\vec{v}$, pressure field $\mns$, left Cauchy--Green tensor field $\lcgnc$ and the temperature field $\temp$.

\subsubsection{Remarks}
\label{sec:remarks} Let us now briefly comment on the derivation of the model and some of its features. First, if $\mu$ is a constant or a slowly varying function of the temperature, and if $\lcgnc$ is close to the identity tensor, then \eqref{eq:77} reduces to the standard heat equation with the source term corresponding to the ``solvent'' viscosity,
\begin{equation}
  \label{eq:80}
  \rho \sheatcapvol^{\mathrm{iNSE}} \dd{\temp}{t}
  =
  \divergence \left(\kappa(\temp) \nabla \temp \right) + 2 \nu(\temp) \tensordot{\traceless{\gradsym}}{\traceless{\gradsym}}.
\end{equation}
(The proximity of $\lcgnc$ to the identity tensor means that the elastic part of the deformation is small.) In this sense, the standard evolution equation for the temperature~\eqref{eq:80} can be seen as an approximation of the exact evolution equation~\eqref{eq:77}. 

Second, most of the non-standard terms in the temperature evolution equation arise due to the dependence of $\mu$ on the temperature. On the other hand, the impact of the temperature dependent ``solvent'' viscosity $\nu$ on the structure of the equations is minimal. It in principle suffices to replace constant $\nu$ in the classical equations by the temperature dependent~$\nu(\temp)$. Since $\mu$ characterises the elastic response, see Section~\ref{sec:ansatz-free-energy}, the temperature dependent $\mu$ is tantamount to the temperature dependent heights of the jumps in the creep and stress relaxation tests. (See~\cite{rehor.m.pusa.v.ea:on} for the analysis of creep and stress relaxation experiments for non-linear viscoelastic materials.) This provides one a quick experimentally oriented characterisation of a class of viscoelastic materials that must be very carefully treated with respect to the temperature changes. If the given material exhibits strong temperature dependence of the jump heights in creep or stress relaxation tests, then working with~\eqref{eq:80} instead of~\eqref{eq:77} will very likely lead to inaccurate results.

Third, the elastic contribution in the free energy \emph{ansatz} can be easily modified. There is no need to consider the compressible neo-Hookean response. Other non-linear incompressible/compressible elastic models can be used as well. (See for example~\cite{horgan.co.saccomandi.g:constitutive} or \cite{horgan.co.murphy.jg:constitutive} and references therein for a list of stored energies used in the elasticity.) This would naturally lead to various generalisations of the standard Maxwell/Oldroyd-B model and it would expand the parameter space for the fit of specific experimental data. 

Fourth, another degree of complexity can be added to the constitutive relations provided that the underlying kinematics is chosen to be more involved than that depicted in Figure~\ref{fig:viscoelastic-kinematics}. In particular, the kinematics can be constructed in such a way that it reflects \emph{multiple} spring-dashpot models for viscoelastic materials. For example, \emph{isothermal} Burgers type models have also been constructed using the procedure, see~\cite{karra.s.rajagopal.kr:thermodynamic,karra.s.rajagopal.kr:development} or \cite{malek.j.rajagopal.kr.ea:thermodynamically}. Generalisations of these models to the \emph{non-isothermal} setting can be sought for by appealing to the same concepts as above.

Fifth, the treatment of the incompressibility can be made more formal using the concept of Lagrange multiplier. (The incompressibility constraint is enforced by a Lagrange multiplier technique, and the ``pressure'' is then shown to be related to the Lagrange multiplier.) This approach is described in detail in~\cite{malek.j.rajagopal.kr.ea:on}. Here we have adopted, for the sake of simplicity, a less formal treatment. Nevertheless the Lagrange multiplier approach would lead to the same model.

\section{Semianalytical solutions of governing equations}
\label{sec:analyt-solut-govern}
Let us now find analytical solutions of the governing equations in some typical settings frequently encountered in rheology. First, see Section~\ref{sec:cylindr-couette-flow-1}, we will be interested in the steady cylindrical Couette flow with heated/thermally isolated wall. As shown below, this problem posses an analytical solution for the temperature, velocity and left Cauchy--Green fields. However, the fact that we are dealing with the steady flow in a special geometry results in the cancellation of the majority of the terms in the temperature equation~\eqref{eq:77}. While the resulting explicit analytical solution is still useful in testing of numerical schemes, it does not provide a particularly deep insight into the impact of the additional terms in~\eqref{eq:77} to the flow dynamics. 

A different flow problem is necessary to fully analyse the impact of the additional terms in~\eqref{eq:77}. It turns out that unsteady biaxial extension problem can serve this purpose. This problem is solved in Section~\ref{sec:time-depend-sque}, and the semianalytical solution is found via a numerical integration of a simple system of ordinary differential equations.

For the sake of simplicity, in both cases we assume that only the coefficient~$\mu$ depends on the temperature, and that the other material coefficients $\nu$, $\nu_1$, $\kappa$ and $\cheatvol^{\mathrm{iNSE}}$ are constant. In particular, we consider exponential dependence of $\mu$ on the temperature, 
\begin{equation}
  \label{eq:87}
  \mu(\temp) = \tilde{\mu}_{\reference} \exponential{\alpha(\temp - \temp_{\reference})}.
\end{equation}

Note that if the specific heat at constant volume $\cheatvol$, see~\eqref{eq:42}, is a constant, then a simple integration in~\eqref{eq:39} yields the following formula for the entropy
\begin{equation}
  \label{eq:121}
  \entropy(\temp, \lcgnc) 
  = 
  \cheatvol^{\mathrm{iNSE}} \log \left( \frac{\temp}{\temp_{\reference}} \right) 
  - 
  \frac{1}{2\rho} \dd{\mu}{\temp} 
  \left(
    \Tr \lcgnc
    -
    3
    -
    \ln \det \lcgnc
  \right)
  .
\end{equation}
Substituting~\eqref{eq:87} into~\eqref{eq:121} then yields an explicit formula for the entropy in terms of $\temp$ and $\lcgnc$. The entropy is given up to an additive constant, which is expected in a phenomenological type theory. Having the explicit formula for the entropy, we will be able to quantitatively document the consequences of the second law of thermodynamics.  

\subsection{Steady cylindrical Couette flow}
\label{sec:cylindr-couette-flow-1}
\begin{figure}[h]
  \centering
  \subfloat[\label{fig:couette-flow}Cylindrical Couette flow.]{\includegraphics[width=0.3\textwidth]{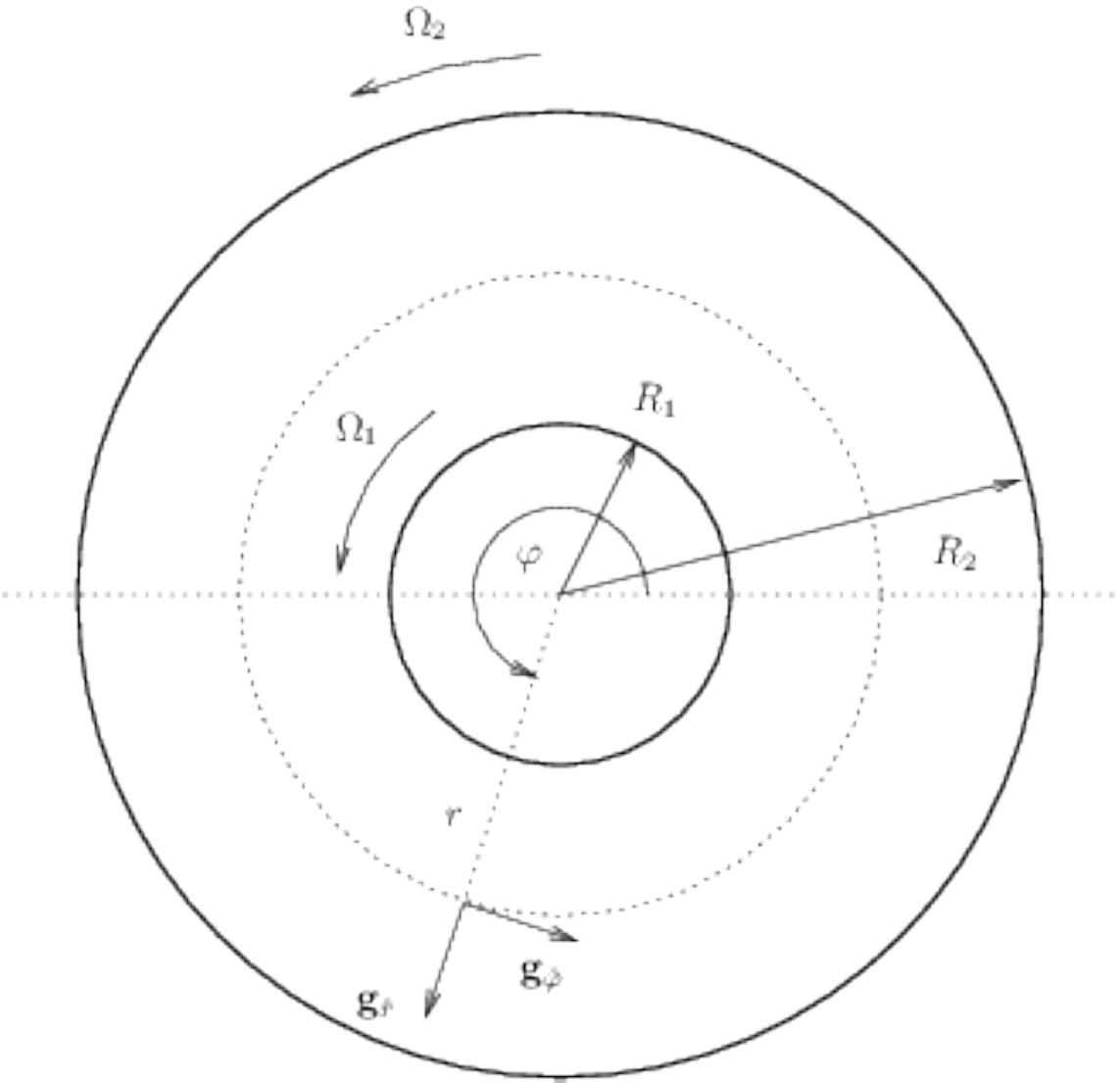}}
  \qquad
  \subfloat[\label{fig:biaxial}Biaxial extension.]{\includegraphics[width=0.3\textwidth]{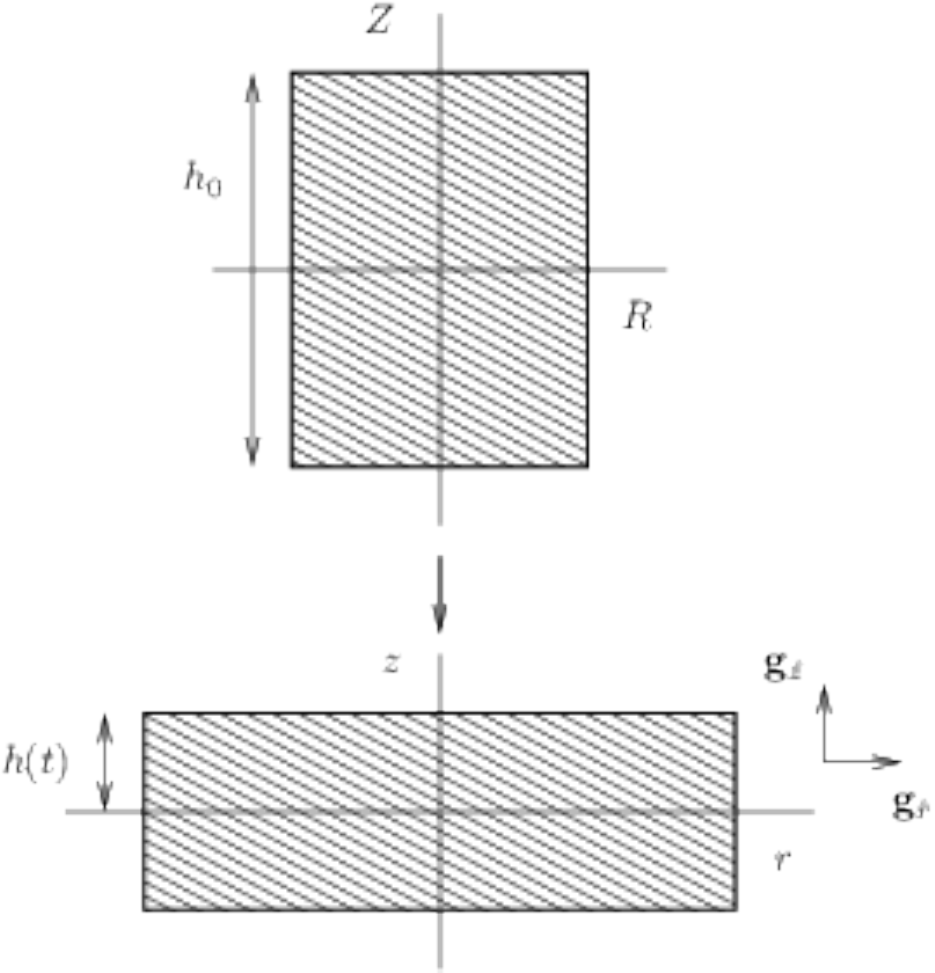}}
  \caption{Problem geometry.}
\end{figure}

Let us now investigate the steady flow of a viscoelastic fluid with temperature dependent material coefficients in the standard Couette flow geometry, see Figure~\ref{fig:couette-flow}. The fluid of interest is placed in between two infinite concentric cylinders of radii $R_1$ and $R_2$, $R_1<R_2$. The cylinders are rotating with the angular velocities  $\Omega_1$ (inner cylinder) and $\Omega_2$ (outer cylinder). The inner cylinder is assumed to be kept at the fixed temperature $\temp_1$, while the other cylinder is thermally isolated (no heat flux through the wall). Consequently, if the no-slip boundary condition is assumed at the walls, we get the following boundary conditions for the temperature and velocity field,
\begin{subequations}
  \label{eq:boundary-conditions}
  \begin{align}
    \label{eq:81}
    \left. \vec{v} \right|_{r=R_1} &= R_1 \Omega_1 \cobvec{\hatp}, \\
    \label{eq:79}
    \left. \vec{v} \right|_{r=R_2} &= R_2 \Omega_2 \cobvec{\hatp}, \\
    \label{eq:82}
    \left. \temp \right|_{r=R_1} &= \temp_1, \\
    \label{eq:83}
    \left. \vectordot{\left(\nabla \temp\right)}{\cobvec{\hatr}} \right|_{r=R_2} &= 0.
  \end{align}
\end{subequations}
Here $\cobvec{\hatp}$ and $\cobvec{\hatr}$ denote the normed basis vectors in the cylindrical coordinate system, see Figure~\ref{fig:couette-flow}. 

Since the cylinders are infinite, and the problem has the rotational symmetry, we shall seek the velocity, temperature and pressure field in the form
\begin{equation}
  \label{eq:1415}
  \vec{v}
  =
  \vhatp(r) \cobvec{\hatp}
  ,
  \qquad
  \temp = \temp(r), 
  \qquad
  \mns = \mns(r),
\end{equation}
while the left Cauchy--Green tensor representing the response from the natural to the current configuration is assumed to take the form
\begin{equation}
  \label{eq:1412}
  \lcgnc
  =
  \begin{bmatrix}
    \Brrhat(r) & \Brphat(r) & 0 \\
    \Bprhat(r) & \Bpphat(r) & 0 \\
    0 & 0 & \Bzzhat(r)
  \end{bmatrix}
  .
\end{equation}
Note that the chosen \emph{ansatz} for the velocity field automatically satisfies~\eqref{eq:74}. 
The assumptions lead to the following expressions for the velocity gradient, the symmetric part of the velocity gradient, the convective term, the divergence of~$\lcgnc$, and the upper convected derivative of $\lcgnc$,
\begin{subequations}
  \label{eq:formulae-cylindrical}
  \begin{align}
    \label{eq:88}
    \nabla \vec{v}
    &=
    \begin{bmatrix}
      0 & -\omega & 0 \\
      r \dd{\omega}{r}  + \omega & 0 & 0 \\
      0 & 0 & 0
    \end{bmatrix}
    ,
    \\
    \label{eq:90}
    \gradsym
    &=
    \begin{bmatrix}
      0 & \frac{r}{2} \dd{\omega}{r} & 0 \\
      \frac{r}{2} \dd{\omega}{r} & 0 & 0 \\
      0 & 0 & 0
    \end{bmatrix}
    ,
    \\
    \label{eq:91}
    \dd{\vec{v}}{t}
    &=
    \begin{bmatrix}
      -r \omega^2 \\
      0 \\
      0
    \end{bmatrix}
    ,
    \\
    \label{eq:92}
    \divergence \lcgnc
    &=
    \begin{bmatrix}
      \frac{1}{r}
      \dd{}{r}
      \left(
        r \Brrhat
      \right)
      -
      \frac{\Bpphat}{r}
      \\
      \dd{\Bprhat}{r} + \frac{\Bprhat + \Brphat}{r} 
      \\
      0
    \end{bmatrix}
    ,
    \\
    \label{eq:93}
    \fid{\overline{\lcgnc}}
    &=
    \begin{bmatrix}
      0 & -r \dd{\omega}{r} \Brrhat & 0 \\
      -r \dd{\omega}{r} \Brrhat & -2 r \dd{\omega}{r} \Bprhat & 0 \\
      0 & 0 & 0
    \end{bmatrix}
    ,
  \end{align}
\end{subequations}
where we have introduced the angular velocity $\omega(r)$, $\vhatp(r) =_{\bydefinition} \omega(r) r$. Further, it holds $\dd{\temp}{t}=0$. Using~\eqref{eq:formulae-cylindrical}, we see that the governing equations for the velocity field~\eqref{eq:71} reduce, in the absence of body force $\vec{b}$, to
\begin{subequations}
  \label{eq:84}
  \begin{equation}
    \label{eq:85}
    \begin{bmatrix}
      - \rho r \omega^2 \\
      0 \\
      0
    \end{bmatrix}
    =
    \begin{bmatrix}
     \dd{}{r}\left(\mns+\mu \left(\Brrhat-\frac13(\Brrhat+\Bpphat+\Bzzhat) \right)\right)
    +
    \mu
    \frac{\Brrhat - \Bpphat}{r}
      \\
      \frac{1}{r^2}
      \dd{}{r}\left(r^3 \nu \dd{\omega}{r} + \mu(\temp) r^2 \Bprhat \right)
      \\
      0
    \end{bmatrix}
    ,
  \end{equation}
  while the governing equations~\eqref{eq:73} for the left Cauchy--Green tensor $\lcgnc$ read
  \begin{equation}
    \label{eq:86}
    \nu_1   
    \begin{bmatrix}
      0 & -r \dd{\omega}{r} \Brrhat & 0 \\
      -r \dd{\omega}{r} \Brrhat & -2 r \dd{\omega}{r} \Bprhat & 0 \\
      0 & 0 & 0
    \end{bmatrix}
    +
    \mu(\temp)
    \begin{bmatrix}
      \Brrhat -1 & \Brphat & 0 \\
      \Bprhat & \Bpphat -1 & 0 \\
      0 & 0 & \Bzzhat -1
    \end{bmatrix}
    =
    \begin{bmatrix}
      0 \\
      0 \\
      0
    \end{bmatrix}
    .
  \end{equation}
\end{subequations}

The solution to the transport equation for $\lcgnc$ reads
\begin{subequations}
  \label{eq:94}
  \begin{align}
    \label{eq:95}
    \Brrhat &= 1, \\
    \label{eq:96}
    \Bzzhat &= 1, \\
    \label{eq:97}
    \Brphat &=\frac{\nu_1}{\mu(\temp)}r\dd{\omega}{r}, \\
    \label{eq:98}
    \Bpphat &=1 + 2\left( \frac{\nu_1}{\mu(\temp)}r\dd{\omega}{r} \right)^2.
  \end{align}
\end{subequations}
Note that $\Bpphat = 1 + 2 \left(\Brphat\right)^2$. Substituting~\eqref{eq:97} into the second equation in~\eqref{eq:85} then yields the governing equation for the angular velocity $\omega$,
\begin{equation}
  \label{eq:99}
  \frac{1}{r^2}
  \dd{}{r}
  \left(
    r^3 \left(\nu + \nu_1 \right)
    \dd{\omega}{r}
  \right)
  =
  0
  .
\end{equation}
Using the boundary conditions $\left. \omega \right|_{r=R_1} = \Omega_1$ and $\left. \omega \right|_{r=R_2} = \Omega_2$ that follow from the definition of the angular velocity and boundary conditions~\eqref{eq:boundary-conditions}, we get the sought velocity field
\begin{equation}
  \label{eq:100}
  \omega =\frac{R_1^2\Omega_1-R_2^2\Omega_2}{R_1^2-R_2^2}-\frac{R_1^2R_2^2(\Omega_1-\Omega_2)}{r^2(R_1^2-R_2^2)}.
\end{equation}
Note that the velocity field is the same as the velocity field in the classical incompressible Navier--Stokes fluid, see for example~\cite{taylor.gi:stability}.

Let us now proceed with the solution of the equation for the temperature field~\eqref{eq:77}. Since $\dd{\temp}{t}=0$, we see that~\eqref{eq:77} reduces to
\begin{equation}
  \label{eq:101}
  0
  =
  2 \nu \tensordot{\traceless{\gradsym}}{\traceless{\gradsym}}
  +
  \divergence \left(\kappa \nabla \temp \right)
  +
  \temp\dd{\mu}{\temp}
  \tensordot{
    \traceless{\left(\lcgnc\right)}
  }
  {\traceless{\gradsym}}
  +
  \frac{\mu(\temp)}{2 \nu_1}
  \left(
    \mu(\temp) - \temp \dd{\mu}{\temp}(\temp)
  \right)
  \left(
    \Tr \lcgnc + \Tr \left( \inverse{\lcgnc} \right) - 6
  \right)
  .
\end{equation}
Since the left Cauchy--Green tensor reads
\begin{equation}
  \label{eq:102}
  \lcgnc
  =
  \begin{bmatrix}
    1 & \Brphat & 0 \\
    \Brphat & \Bpphat & 0 \\
    0 & 0 & 1
  \end{bmatrix}
\end{equation}
one gets
\begin{equation}
  \label{eq:103}
  \inverse{\lcgnc}
  =
  \begin{bmatrix}
    \frac{\Bpphat}{\Bpphat - \left( \Brphat \right)^2} & -\frac{\Brphat}{\Bpphat - \left( \Brphat \right)^2} & 0 \\
    -\frac{\Bpphat}{\Bpphat - \left( \Brphat \right)^2} & \frac{1}{\Bpphat - \left( \Brphat \right)^2} & 0 \\
    0 & 0 & 1
\end{bmatrix}
,
\end{equation}
which upon using the identity $\Bpphat = 1 + 2 \left(\Brphat\right)^2$ implies that $\Tr \inverse{\lcgnc} = 3$. Consequently, we see that
\begin{subequations}
  \label{eq:104}
  \begin{align}
    \label{eq:105}
    \Tr \lcgnc + \Tr \left( \inverse{\lcgnc} \right) - 6 &= 2 \left( \frac{\nu_1}{\mu(\temp)} r \dd{\omega}{r} \right)^2, \\
    \label{eq:106}
    \tensordot{
      \traceless{\left(\lcgnc\right)}
    }
    {\traceless{\gradsym}}
    &=
    \frac{\nu_1}{\mu(\temp)}
    \left( r \dd{\omega}{r} \right)^2
    .
  \end{align}
\end{subequations}

Using~\eqref{eq:104} in~\eqref{eq:101}, we see that the equation for the temperature reduces to
\begin{equation}
  \label{eq:107}
  0
  =
  \nu
  \left( r \dd{\omega}{r} \right)^2
  +
  \frac{1}{r} \dd{}{r} \left(r\kappa \dd{\temp}{r}\right)
  +
  \nu_1
  \left( r \dd{\omega}{r} \right)^2.
\end{equation}
Since the function $\left( \nu + \nu_1\right) \left( r \dd{\omega}{r} \right)^2$ is a known function, see~\eqref{eq:100}, it remains to solve the linear equation~\eqref{eq:107} for $\temp$. The solution to~\eqref{eq:107} with $\omega$ given by~\eqref{eq:100} reads
\begin{equation}
  \label{eq:108}
  \temp = C_1 - \frac{C}{4r^2} + C_2 \ln r,
\end{equation}
where $C_1$ and $C_2$ are integration constants, and 
\begin{equation}
  \label{eq:9}
  C =_{\bydefinition} \frac{ \nu + \nu_1 }{\kappa} \left( \frac{2 R_1^2 R_2^2 (\Omega_1 - \Omega_2)}{R_1^2 - R_2^2} \right)^2.
\end{equation}
It remains to fix the integration constants $C_1$ and $C_2$ using the boundary conditions for the temperature, see~\eqref{eq:82} and \eqref{eq:83}. The latter boundary condition translates, in virtue of the structure of the temperature field, to $\left. \dd{\temp}{r} \right|_{r=R_2} = 0$. The complete solution for the temperature field then reads
\begin{equation}
  \theta=\theta_1 +\frac{C}{4}\left(\frac{1}{R_1^2}-\frac{1}{r^2}-\frac{2}{R_2^2}\ln\frac{r}{R_1}\right).
\end{equation}
Note that once we know the temperature field, the material coefficient $\mu$ in~\eqref{eq:94} is a known function of position, hence~\eqref{eq:94} yields explicit formulae for the components of the tensor field $\lcgnc$. Finally, the pressure $\mns$ is obtained from the first equation in the system~\eqref{eq:85} via a simple integration
\begin{equation}
  \label{eq:1}
  \mns
  =
  \left. \mns \right|_{r=R_1} 
  + 
  \frac{2}{3}
  \left[
    \mu(\temp(r))(\Brphat(r))^2
    -
    \mu(\temp(R_1))(\Brphat(R_1))^2
  \right] 
  + 
  \int_{s=R_1}^r
  \left[
    -\rho\omega^2(s) s
    +
    2\mu\frac{(\Brphat(s))^2}{s}
  \right]\, \diff s,
\end{equation}
where $\left. \mns \right|_{r=R_1}$ denotes the pressure value at $r=R_1$.

\subsubsection{Summary}
\label{sec:summary-2}

The pressure $\mns$, temperature $\temp$, velocity $\vhatp =_{\bydefinition} r \omega(r)$ and left Cauchy--Green field are in the steady cylindrical Couette flow given by the formulae
\begin{subequations}
  \label{eq:solution-cylindrical-couette}
  \begin{align}
    \label{eq:3}
    \omega &=\frac{R_1^2\Omega_1-R_2^2\Omega_2}{R_1^2-R_2^2}-\frac{R_1^2R_2^2(\Omega_1-\Omega_2)}{r^2(R_1^2-R_2^2)}, \\
    \label{eq:2}
    \theta &=\theta_1 +\frac{C}{4}\left(\frac{1}{R_1^2}-\frac{1}{r^2}-\frac{2}{R_2^2}\ln\frac{r}{R_1}\right), \\
    \label{eq:4}
    \mns &= 
    \left. \mns \right|_{r=R_1} 
    + 
    \frac{2}{3}
    \left[
      \mu(\temp(r))(\Brphat(r))^2
      -
      \mu(\temp(R_1))(\Brphat(R_1))^2
    \right] 
    + 
    \int_{s=R_1}^r
    \left[
      -\rho\omega^2(s) s
      +
      2\mu\frac{(\Brphat(s))^2}{s}
    \right]\, \diff s,
    \\
    \label{eq:5}
    \Brrhat &= 1, \\
    \label{eq:6}
    \Bzzhat &= 1, \\
    \label{eq:7}
    \Brphat &=\frac{\nu_1}{\mu(\temp)} \frac{2 R_1^2 R_2^2 (\Omega_1 - \Omega_2)}{r^2 \left(R_1^2 - R_2^2\right)}, \\ 
    \label{eq:8}
    \Bpphat &= 1 + 2 \left(\Brphat\right)^2.
  \end{align}
\end{subequations}
As already noted, the steady velocity profile is the same as for the standard incompressible Navier--Stokes fluid. Concerning the temperature field, it is useful to compare the predictions based on the naive temperature evolution equation~\eqref{eq:80} and the complete temperature evolution equation~\eqref{eq:77}. If the temperature field was calculated using the primitive temperature evolution equation~\eqref{eq:80}, then the temperature field would be qualitatively the same. Formula~\eqref{eq:2} would be still valid, but the value of the constant $C$ would be given as
\begin{equation}
  \label{eq:10}
  C_{\mathrm{naive}} =_{\bydefinition}  \frac{\nu}{\kappa} \left( \frac{2 R_1^2 R_2^2 (\Omega_1 - \Omega_2)}{R_1^2 - R_2^2} \right)^2
\end{equation}
instead of~\eqref{eq:9}. The difference between the predicted temperature values would further induce differences in the predicted left Cauchy--Green tensor field and the pressure field.

\subsection{Time dependent biaxial extension}
\label{sec:time-depend-sque}

The biaxial extension is a deformation that can be achieved for example in the lubricated squeeze flow, see~\cite{chatraei.s.macosko.cw.ea:lubricated}, \cite{kompani.m.venerus.cd:equibiaxial}, \cite{nasseri.s.bilston.l.ea:modelling}, \cite{rehor.m.prusa.v:squeeze} or \cite{rehor.m.pusa.v.ea:on}. It takes place in a sample that undergoes the deformation shown in Figure~\ref{fig:biaxial}. The cylindrical sample of initial height $h_0$ and radius $R_0$ is deformed to a cylindrical body of height $h(t)$, and it is assumed that the horizontal planes remain horizontal planes during the deformation. Function $h(t)$ is the given datum and the task is to find the velocity, pressure, temperature and the left Cauchy--Green fields in the sample. 

The sample is assumed to be thermally isolated, that is $\vectordot{\nabla \temp}{\vec{n}} = 0$ holds on the boundary of the sample. The boundary conditions for the velocity are fixed in such a way that the biaxial extension automatically fulfills the boundary conditions, see, for example, \cite{engmann.j.servais.c.ea:squeeze} or \cite{rehor.m.pusa.v.ea:on} for details. The sample is assumed to be initially in equilibrium, which translates to the initial conditions
\begin{equation}
  \label{eq:110}
  \left. \vec{v} \right|_{t=0} = \vec{0}, \qquad \left. \lcgnc \right|_{t=0} = \identity, \qquad \left. \temp \right|_{t=0} = \temp_{\reference},
\end{equation}
where $\temp_{\reference}$ is a given reference temperature.

If the cylindrical coordinate system is used, then the deformation from the reference $\vec{X} = [R, \Phi, Z]$ to the current configuration $\vec{x} = [r,\varphi,z]$ is described as
\begin{subequations}
  \label{eq:motion}  
  \begin{align}
    \label{eq:115}
    r &= \sqrt{\frac{h_0}{h}} R, \\
    \label{eq:116}
    \varphi &= \Phi, \\
    \label{eq:117}
    z &= \frac{h}{h_0} Z.
  \end{align}
\end{subequations}
Consequently, the Eulerian velocity field is given as 
$
\vec{v} 
= 
\transpose{
  \begin{bmatrix} 
    -\frac{\dhencky}{2} r &
    0 &
    \dhencky z
  \end{bmatrix}
}
$, where $\dhencky =_{\bydefinition} \pd{\hencky}{t}$ and
\begin{equation}
  \label{eq:12}
  \hencky =_{\bydefinition} \ln \frac{h}{h_0}
\end{equation}
denotes the Hencky strain. This leads to the following expressions for the velocity gradient, the symmetric part of the velocity gradient and the material time derivative of the velocity field
\begin{subequations}
  \label{eq:11}
  \begin{align}
    \label{eq:velocity-gradient}
    \gradvl 
    &= 
    \gradsym
    =
    \begin{bmatrix}
      -\frac{\dhencky}{2} & 0 & 0 \\
      0 & -\frac{\dhencky}{2} & 0 \\
      0 & 0 & \dhencky
    \end{bmatrix}
    ,
    \\
    \label{eq:material-time-derivative}
    \dd{\vecv}{t}
    &=
    \pd{\vecv}{t}
    +
    (
    \vectordot{\vecv}{\nabla}
    )
    \vecv
    =
    \begin{bmatrix}
      -\frac{1}{2} \ddot{\hencky} r + \left( \frac{\dot{\hencky}}{2} \right)^2 r\\
      0\\
      \ddot{\hencky} z+(\dot{\hencky})^2 z
    \end{bmatrix}
    ,
  \end{align}
\end{subequations}
hence the chosen \emph{ansatz} for the deformation indeed leads to a volume preserving deformation.
 
If the left Cauchy--Green tensor $\lcgnc$ is assumed to take the form
\begin{equation}
  \label{eq:13}
  \lcgnc
  =
  \begin{bmatrix}
    \Brrhat & 0 & 0 \\
    0 & \Brrhat & 0 \\
    0 & 0 & \Bzzhat
  \end{bmatrix}
  ,
\end{equation}
where the components $\Brrhat$ and $\Bzzhat$ depend on the time only, then
\begin{subequations}
  \label{eq:14}
  \begin{align}
    \label{eq:15}
    \divergence \lcgnc &= \vec{0}, \\
    \label{eq:16}
    \dd{\lcgnc}{t} &= \pd{\lcgnc}{t}, \\
    \label{eq:20}
    \fid{\overline{\lcgnc}} &=
    \begin{bmatrix}
      \pd{\Brrhat}{t} + \dhencky \Brrhat & 0 & 0 \\
      0 & \pd{\Brrhat}{t} + \dhencky \Brrhat & 0 \\
      0 & 0 & \pd{\Bzzhat}{t} - 2\dhencky \Bzzhat
    \end{bmatrix}
    .
  \end{align}
\end{subequations}
Note that the \emph{ansatz} for the $\lcgnc$ field is constructed in such a way that the $\hat{r}\hat{r}$ and $\hat{\varphi}\hat{\varphi}$ components of $\lcgnc$ are identical. Further, the homogeneity of the $\lcgnc$ and $\gradsym$ fields indicates that the temperature field governed by~\eqref{eq:77} is also homogeneous in space. If the temperature field is homogeneous, $\temp(\vec{x},t)=\temp(t)$, then the material time derivative of the temperature field reads
\begin{equation}
  \label{eq:21}
  \dd{\temp}{t} = \pd{\temp}{t}.
\end{equation}

Using~\eqref{eq:11}, \eqref{eq:13}, \eqref{eq:14} and \eqref{eq:21} in the governing equations~\eqref{eq:maxwell-oldroyd-temperature-dependent-summary} leads to the following set of non-trivial differential equations
\begin{subequations}
  \label{eq:governing-equations}
  \begin{align}
    \label{eq:109}
    \rho \left(-\frac{\ddhencky}{2} + \left(\frac{\dhencky}{2}\right)^2 \right) r &= \pd{\mns}{r}, \\
    \label{eq:111}
    \rho \left(\ddhencky + \left(\dhencky\right)^2 \right) z &= \pd{\mns}{z}, \\
    \label{eq:112}
    \nu_1 \left( \pd{\Brrhat}{t} + \dhencky \Brrhat \right) + \mu(\temp) \left(\Brrhat -1\right) &= 0, \\
    \label{eq:113}
    \nu_1 \left( \pd{\Bzzhat}{t} - 2\dhencky \Bzzhat \right) + \mu(\temp) \left(\Bzzhat -1\right) &=0, 
  \end{align}
and
\begin{multline}
  \label{eq:114}
    \left[
      \rho \sheatcapvol^{\mathrm{iNSE}}
      -
      \left[
        \frac{\temp}{2}
        \ddd{\mu}{\temp}(\temp)
        \left(
          2 \Brrhat + \Bzzhat - 3 - \ln\left( \left(\Brrhat\right)^2 \Bzzhat \right) 
        \right)
      \right]
    \right]
    \pd{\temp}{t}
    =
    3 \nu \left(\dhencky\right)^2 
    + 
    \temp \dd{\mu}{\temp}(\temp) \dhencky \left(\Bzzhat - \Brrhat\right) 
    \\
    + 
    \frac{\mu(\temp)}{2 \nu_1}
    \left(
      \mu(\temp)
      -
      \temp
      \dd{\mu}{\temp}(\temp)
    \right)
    \left[
      2
      \left(
        \Brrhat + \frac{1}{\Brrhat}
      \right)
      +
      \left(
        \Bzzhat + \frac{1}{\Bzzhat}
      \right)
      -
      6
    \right]
    .
\end{multline}
\end{subequations}

Given the function $h(t)$, one can find the Hencky strain $\hencky$, see~\eqref{eq:12}, hence the left hand side of~\eqref{eq:109} and \eqref{eq:111} is a known function. Partial differential equations~\eqref{eq:109} and \eqref{eq:111} are easy to integrate, which yields the formula for the mean normal stress~$\mns$,
\begin{equation}
  \label{eq:89}
  \mns
  = 
  \rho  \left(-\frac{\ddhencky}{2} + \left(\frac{\dhencky}{2}\right)^2 \right) \frac{r^2}{2} 
  +
  \rho \left(\ddhencky + \left(\dhencky\right)^2 \right) \frac{z^2}{2}
  +
  \mns_{\reference},
\end{equation}
where $\mns_{\reference}$ is a constant. The remaining part of the system, that is equations~\eqref{eq:112}, \eqref{eq:113} and \eqref{eq:114} form a system of ordinary differential equations for the temperature and the components of the left Cauchy--Green tensor. This system is subject to the initial conditions $\left. \Brrhat \right|_{t=0} = 1$, $\left. \Bzzhat \right|_{t=0} = 1$  and $\left. \temp \right|_{t=0} = \temp_{\reference}$, and it is easy to solve numerically. 

\begin{figure}[h]
  \centering
  \includegraphics[width=0.4\textwidth]{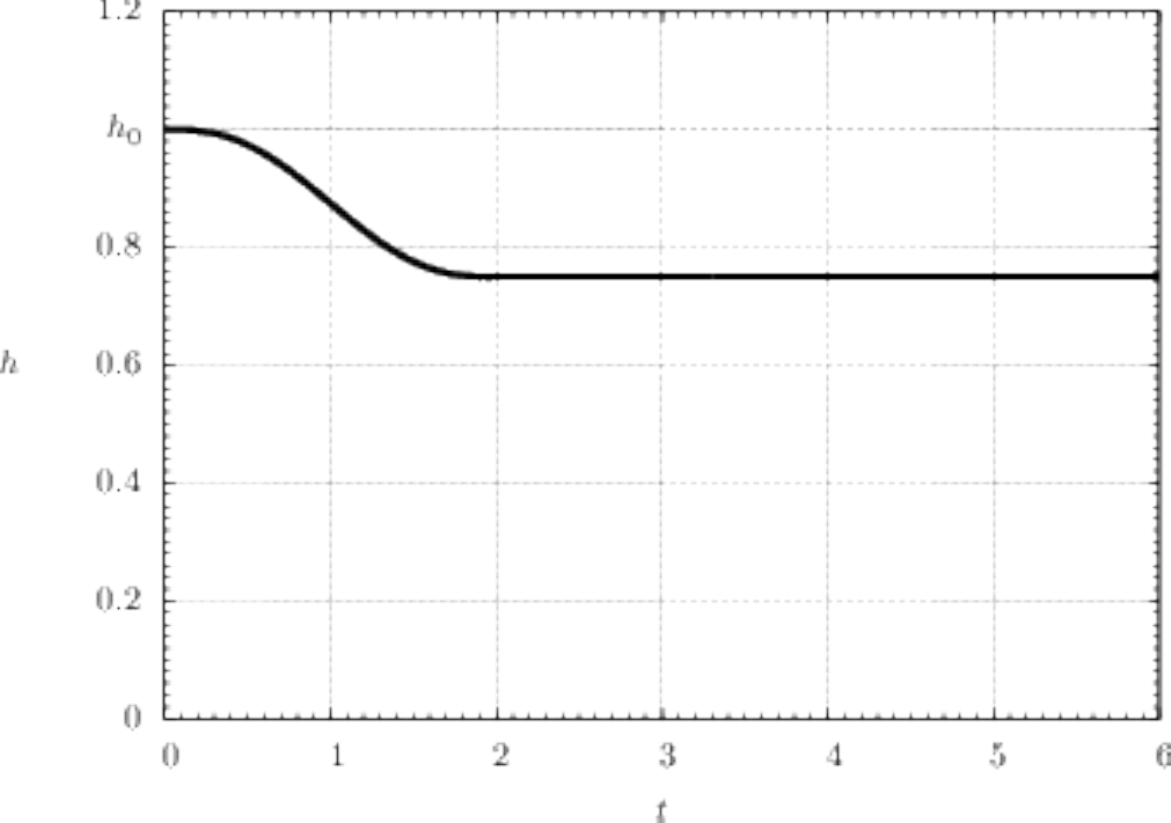}
  \caption{Sample height, smooth transition from the initial height $h_0$ to the final height $h_{\mathrm{end}}$ in the time interval $t_{\mathrm{trans}}$, 
    $h(t)=_{\bydefinition}
    h_0 
    - 
    \left( h_0 -h_{\mathrm{end}} \right) 
    \left(\frac{t}{t_{\mathrm{trans}}}\right)^3 
    \left(
      6 \left(\frac{t}{t_{\mathrm{trans}}}\right)^2
      -
      15
      \left(\frac{t}{t_{\mathrm{trans}}}\right)
      +
      10
    \right)
    $. Parameter values are $h_0=1$, $h_{\mathrm{end}}=\frac{3}{4} h_0$ and $t_{\mathrm{trans}}=2$.}
  \label{fig:sample-height}
\end{figure}

The outputs of numerical simulations are show in Figure~\ref{fig:response-a} and Figure~\ref{fig:response-b}. In both cases we study the response of the fluid subject to the biaxial extension flow generated by the height function shown in Figure~\ref{fig:sample-height}. This height function describes a smooth transition from the initial height $h_0$ to the final height $h_{\mathrm{end}}$, which is chosen as $h_{\mathrm{end}} = \frac{3}{4} h_0$. The figures then show the response of the fluid for various values of the coefficient $\alpha$ in the material function $\mu=\mu(\temp)$, see~\eqref{eq:87}.

The parameter values used in the numerical simulations are given as dimensionless values. This in fact means that we are working with the governing equations as if they were formulated in dimensionless variables. The details of the laborious conversion of the governing equations to their dimensionless form are however not given, since they are irrelevant with respect to the aim of the numerical simulations.
 
\subsubsection{Small solvent viscosity}
\label{sec:small-solv-visc}
\begin{figure}[h]
  \centering
  \subfloat[\label{fig:response-a-entropy} Entropy $\entropy$.]{\includegraphics[width=0.4\textwidth]{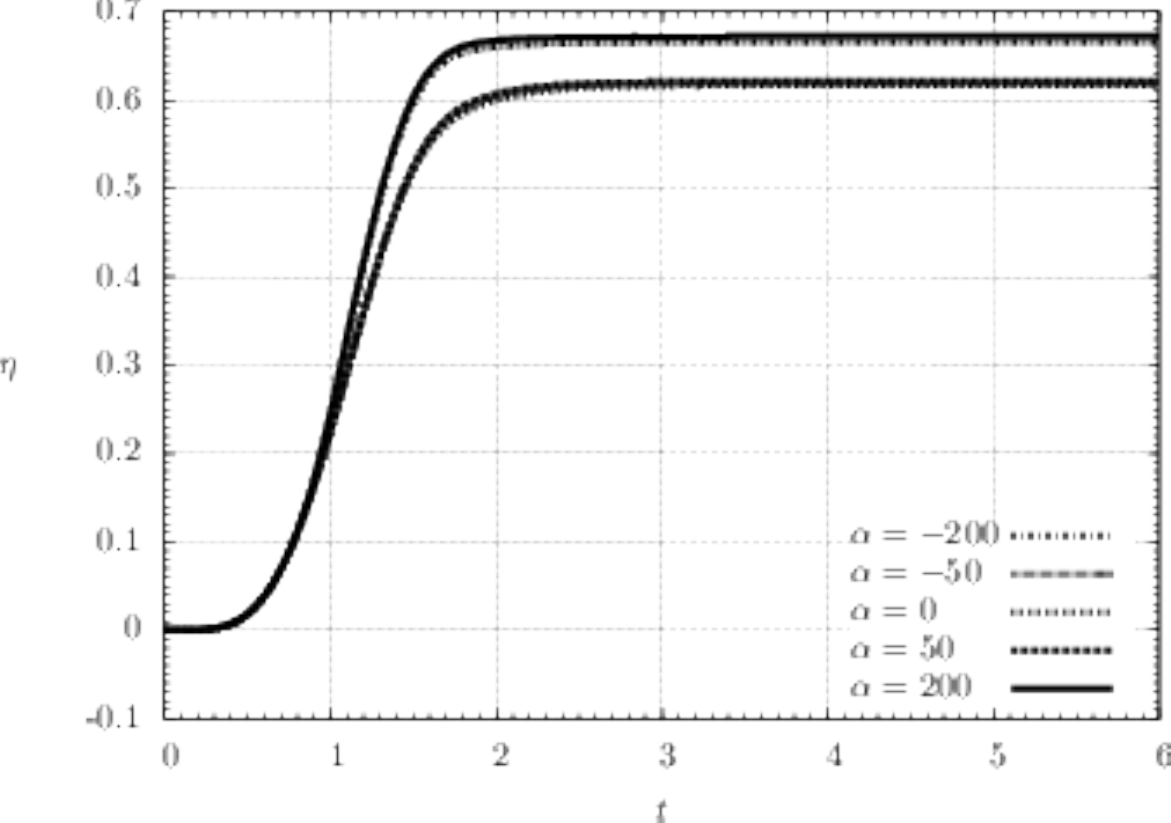}}
  \qquad
  \subfloat[\label{fig:response-a-temperature} Temperature $\temp$.]{\includegraphics[width=0.4\textwidth]{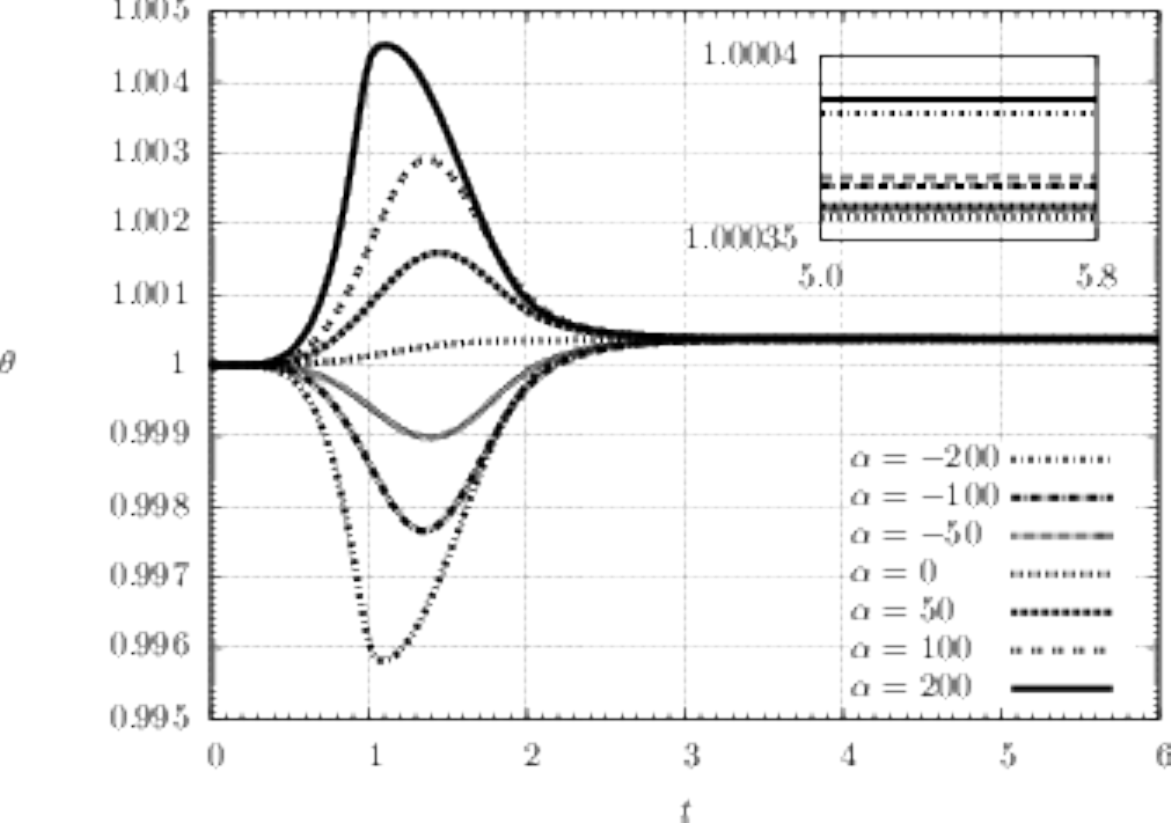}}
  \\
  \subfloat[\label{fig:response-a-lcgncBrr} Left Cauchy--Green tensor $\lcgnc$, component $\Brrhat$.]{\includegraphics[width=0.4\textwidth]{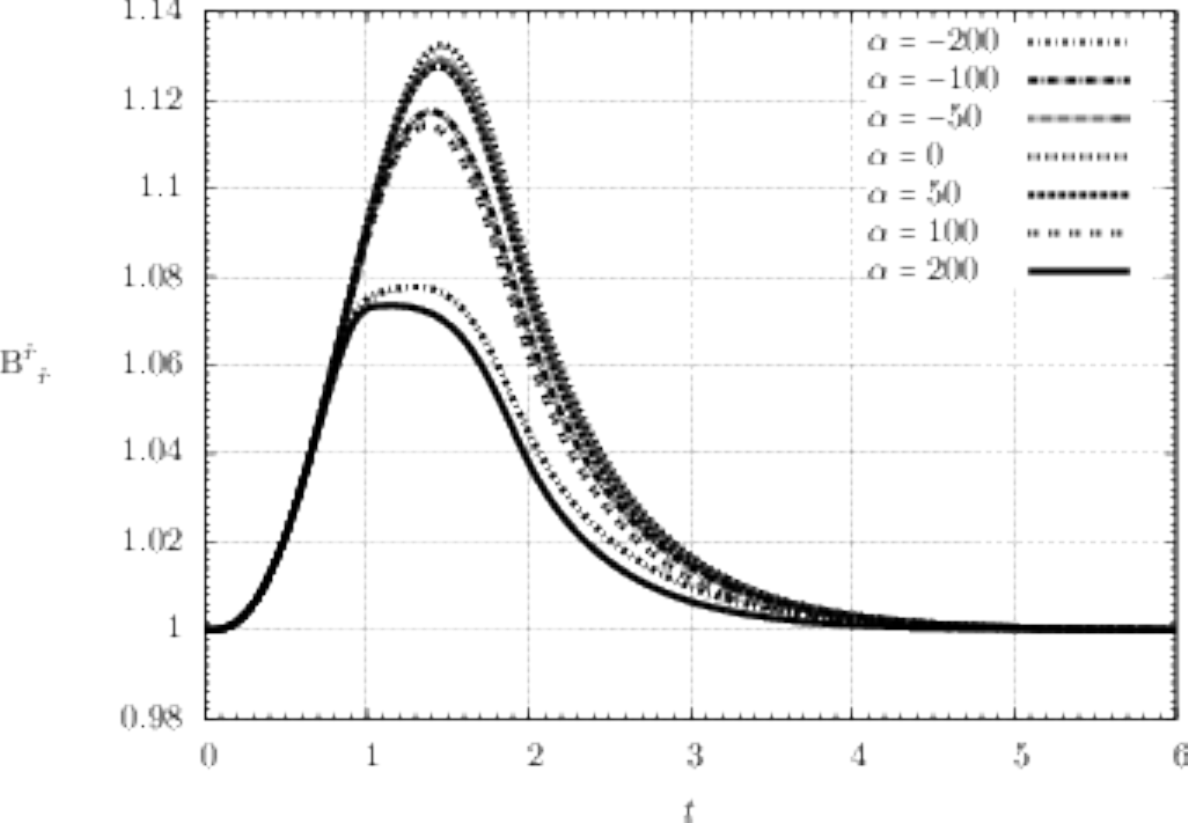}}
  \qquad
  \subfloat[\label{fig:response-a-lcgncBzz} Left Cauchy--Green tensor $\lcgnc$, component $\Bzzhat$.]{\includegraphics[width=0.4\textwidth]{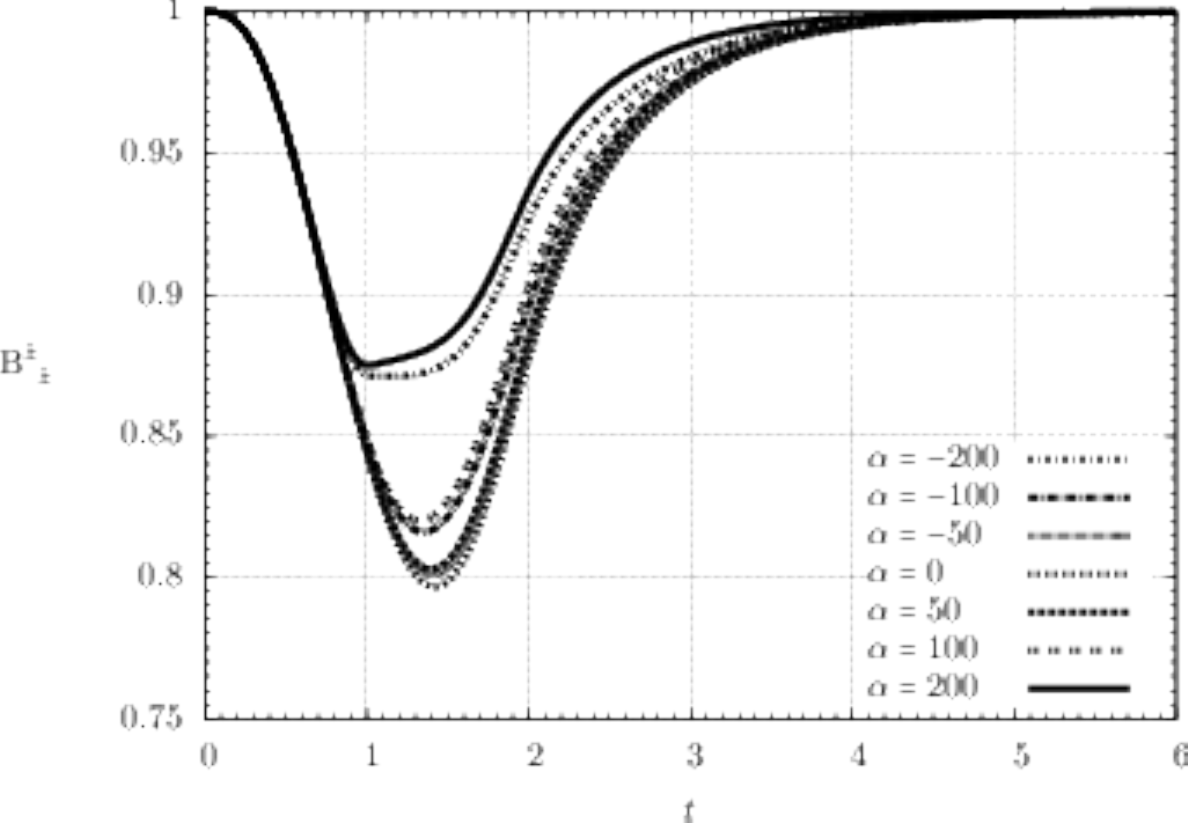}}
  \caption{Time evolution of $\temp$, $\Brrhat$, $\Bzzhat$ and $\entropy$ fields in the sample subject to biaxial extension induced by $h(t)$ shown in Figure~\ref{fig:sample-height}. Parameter values are $\temp_{\reference}=1$, $\mu_{\reference}=\sqrt{5}$, $\nu_1=\sqrt{2}$, $\nu = \sqrt{7}$, $\rho=1$ and $\cheatvol^{\mathrm{iNSE}}=1000 \sqrt{3}$.}
  \label{fig:response-a}
\end{figure}

In the first case, see Figure~\ref{fig:response-a}, the ``solvent'' viscosity $\nu$ is chosen to be relatively small. Consequently, the corresponding term on the right hand side of~\eqref{eq:114} is comparable to the other terms that depend on $\mu$. (See the caption of Figure~\ref{fig:response-a} for the values of the material coefficients.) In other words, there is a non-negligible ``elastic'' part in the temperature evolution equation, and interesting phenomena are expected to occur. Indeed, as shown in Figure~\ref{fig:response-a-temperature}, the temperature can exhibit a non-monotone behaviour during the transition from the initial to the final state. The non-monotone transitional behaviour of the temperature might seem counter-intuitive. However, such a behaviour is not extremely surprising. In fact, such a behaviour is common in \emph{solid} materials where the tension can induce either cooling or heating of the material, see \cite{joule.jp:on*2} and the related modern treatises. Moreover, the non-monotone transitional behaviour is in perfect agreement with the second law of thermodynamics. Since the sample is thermally isolated, \emph{the entropy should grow in time, which is indeed the case}, see Figure~\ref{fig:response-a-entropy}. (The entropy has been calculated using the explicit formula~\eqref{eq:121}.) This means that the second law holds even if the temperature exhibits the non-monotone behaviour.

Note that the final temperature values differ for various values of $\alpha$, see the inset in Figure~\ref{fig:response-a-temperature}. The differences in the final temperature values are however expected since different amount of work has been done on the sample for various values of~$\alpha$. Further, the left Cauchy--Green tensor field $\lcgnc$ relaxes to the equilibrium value $\lcgnc = \identity$ as expected. The transitional behaviour however strongly depends on the value of the coefficient $\alpha$, see Figure~\ref{fig:response-a-lcgncBrr} and Figure~\ref{fig:response-a-lcgncBzz}.

\begin{figure}[h]
  \centering
  \subfloat[\label{fig:response-c-temperature} Temperature $\temp$. Curves for different values of $\alpha$ are identical. Horizontal and vertical axis range corresponds to that in Figure~\ref{fig:response-a-temperature}.]{\includegraphics[width=0.4\textwidth]{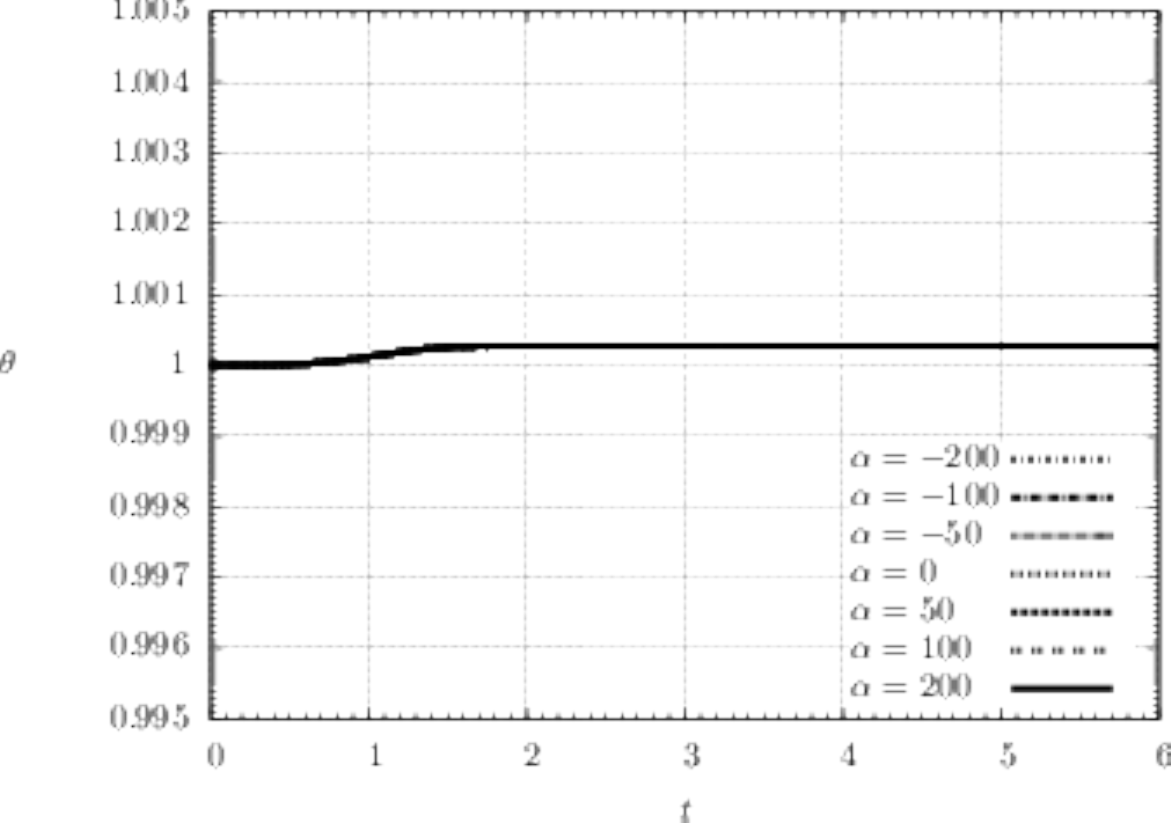}}
  \\
  \subfloat[\label{fig:response-c-lcgncBrr} Left Cauchy--Green tensor $\lcgnc$, component $\Brrhat$.  Horizontal and vertical axis range corresponds to that in Figure~\ref{fig:response-a-lcgncBrr}.]{\includegraphics[width=0.4\textwidth]{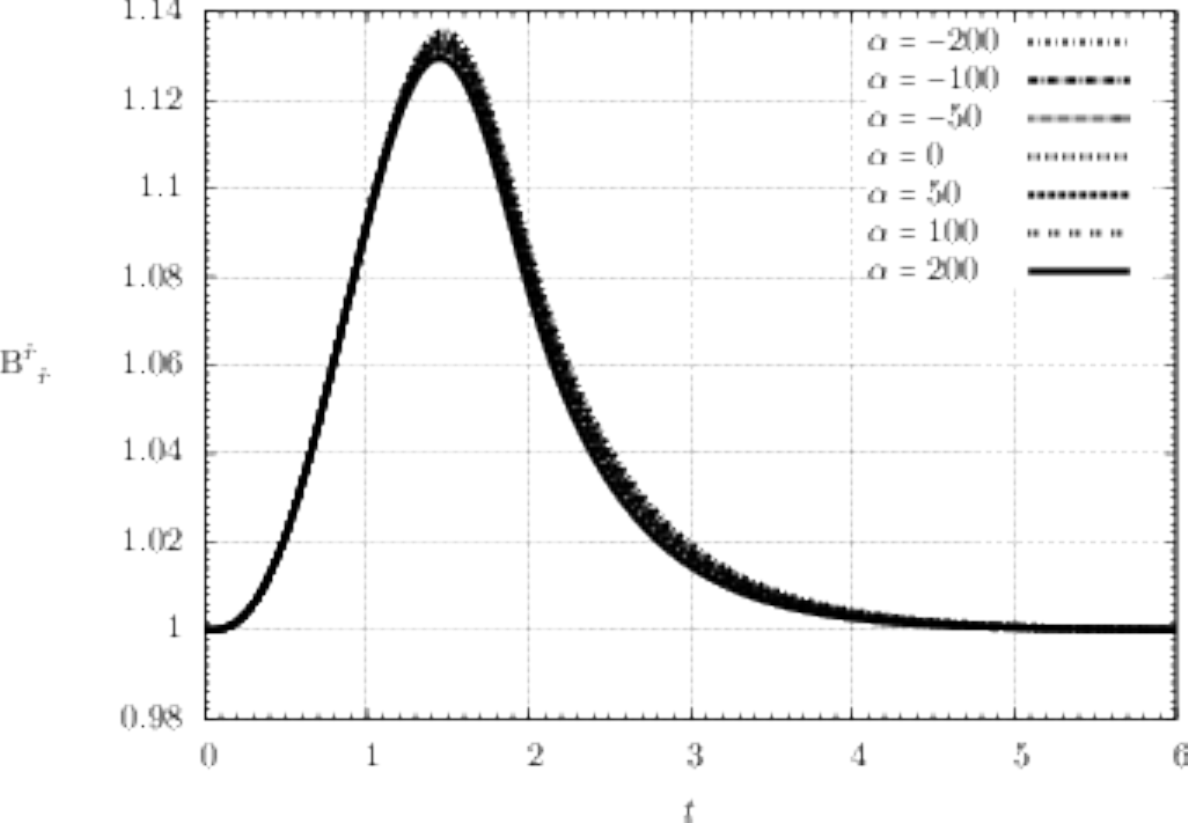}}
  \qquad
  \subfloat[\label{fig:response-c-lcgncBzz} Left Cauchy--Green tensor $\lcgnc$, component $\Bzzhat$.  Horizontal and vertical axis range corresponds to that in Figure~\ref{fig:response-a-lcgncBzz}.]{\includegraphics[width=0.4\textwidth]{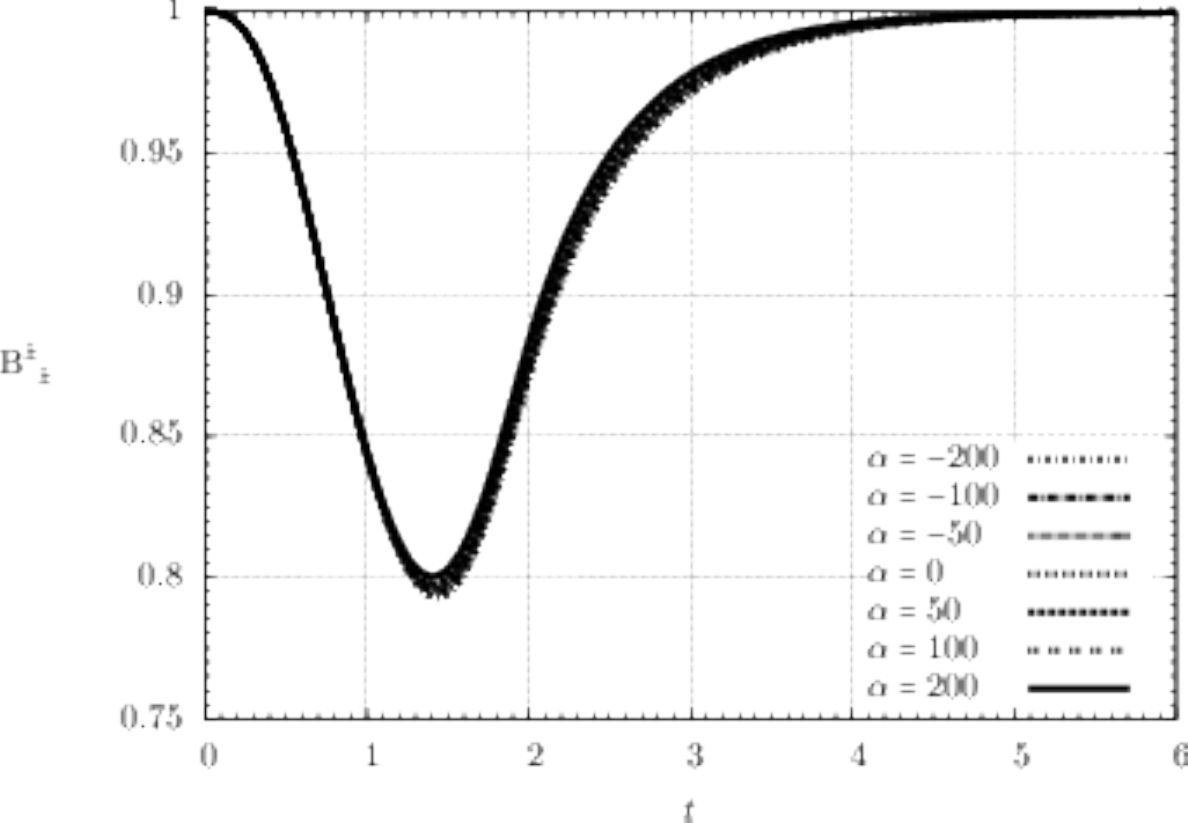}}
  \caption{Time evolution of $\temp$, $\Brrhat$, $\Bzzhat$ and $\entropy$ fields in the sample subject to biaxial extension induced by $h(t)$ shown in Figure~\ref{fig:sample-height}. Parameter values are $\temp_{\reference}=1$, $\mu_{\reference}=\sqrt{5}$, $\nu_1=\sqrt{2}$, $\nu = \sqrt{7}$, $\rho=1$ and $\cheatvol^{\mathrm{iNSE}}=1000 \sqrt{3}$. Exact evolution equation~\eqref{eq:114} replaced by the naive approximation~\eqref{eq:118}.}
  \label{fig:response-c}
\end{figure}

If the exact temperature evolution equation~\eqref{eq:77} was replaced by the naive temperature evolution equation, see~\eqref{eq:80}, then the temperature evolution equation would in our case read
\begin{equation}
  \label{eq:118}
  \rho \sheatcapvol^{\mathrm{iNSE}}
  \pd{\temp}{t}
  =
  3 \nu \left(\dhencky\right)^2 
  .
\end{equation}
Replacing the exact evolution equation~\eqref{eq:114} by~\eqref{eq:118} then leads to the temperature field and left Cauchy--Green field shown in Figure~\ref{fig:response-c}. The temperature values predicted by~\eqref{eq:118} are, except for $\alpha=0$, markedly different from the correct temperature values predicted by~\eqref{eq:114}. Consequently, the predicted values of the left Cauchy--Green tensor are also markedly different from the correct values, compare Figure~\ref{fig:response-c-lcgncBrr} and Figure~\ref{fig:response-c-lcgncBzz} with Figure~\ref{fig:response-a-lcgncBrr} and Figure~\ref{fig:response-a-lcgncBzz}. This mismatch is present irrespective of the fact that the temperature dependent $\mu$ \emph{has been taken into account in the mechanical equations}~\eqref{eq:112} and \eqref{eq:113}.

This observation documents the thesis that the correct formulation of temperature evolution equation can be---for some parameter values and specific processes---of utmost importance. The incorrect description of the energy transfer mechanisms in the fluid of interest can indeed lead to serious errors in the predicted temperature, velocity and stress fields in the fluid. 

\subsubsection{High solvent viscosity}
\label{sec:high-solv-visc-1}
\begin{figure}[h]
  \centering
  \subfloat[\label{fig:response-b-entropy} Entropy $\entropy$.]{\includegraphics[width=0.4\textwidth]{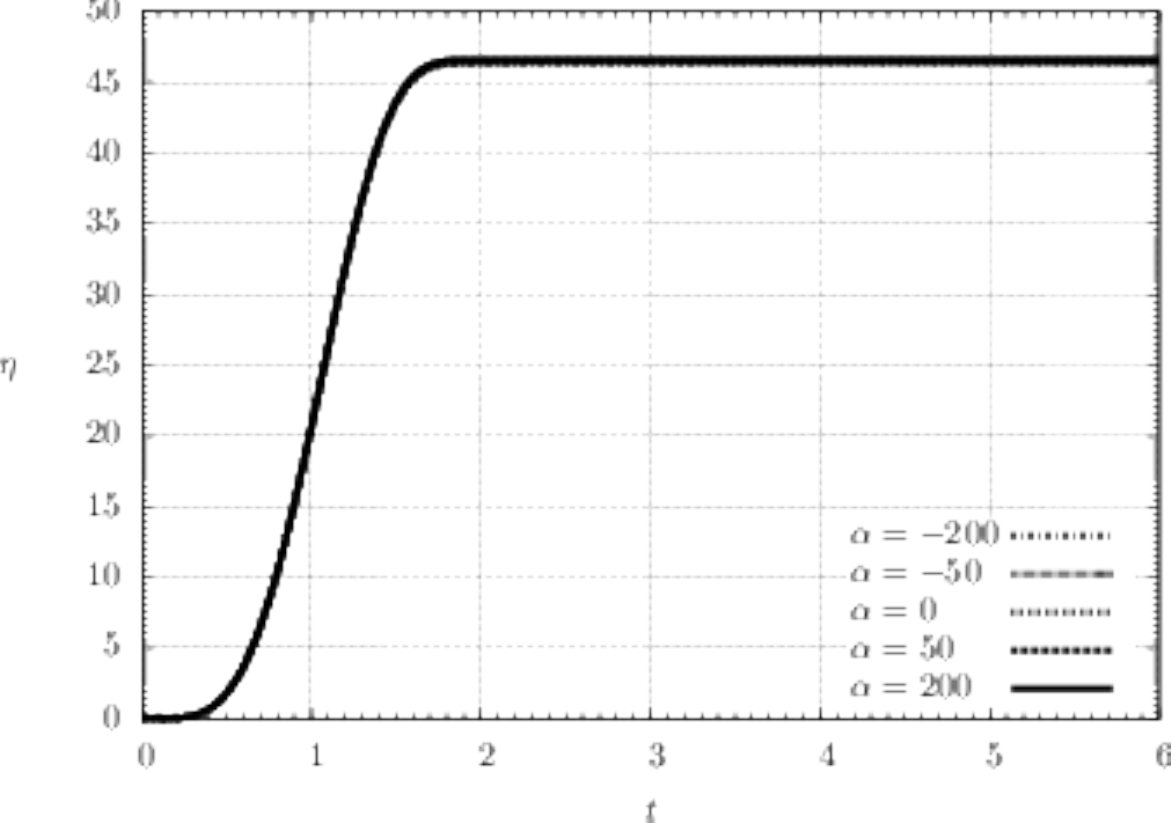}}
  \qquad
  \subfloat[\label{fig:response-b-temperature} Temperature $\temp$.]{\includegraphics[width=0.4\textwidth]{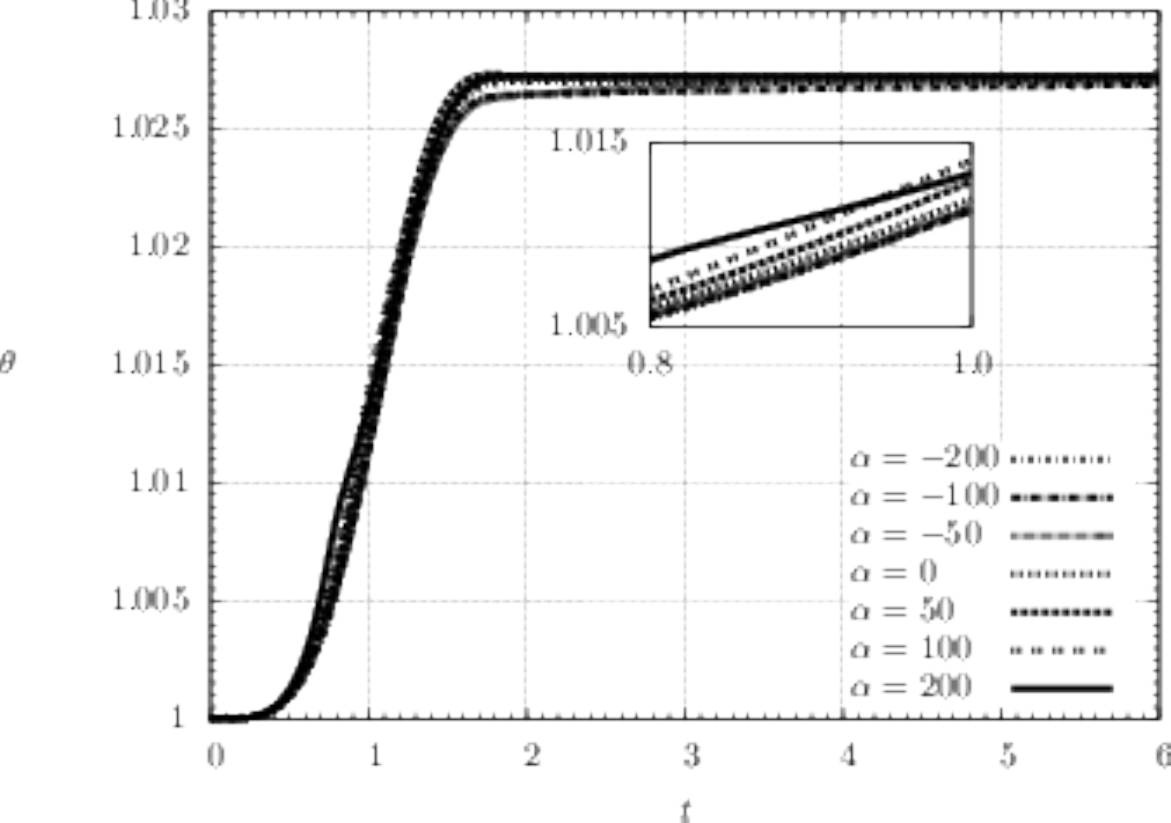}}
  \\
  \subfloat[\label{fig:response-b-lcgncBrr} Left Cacuhy-Green tensor $\lcgnc$, component $\Brrhat$.]{\includegraphics[width=0.4\textwidth]{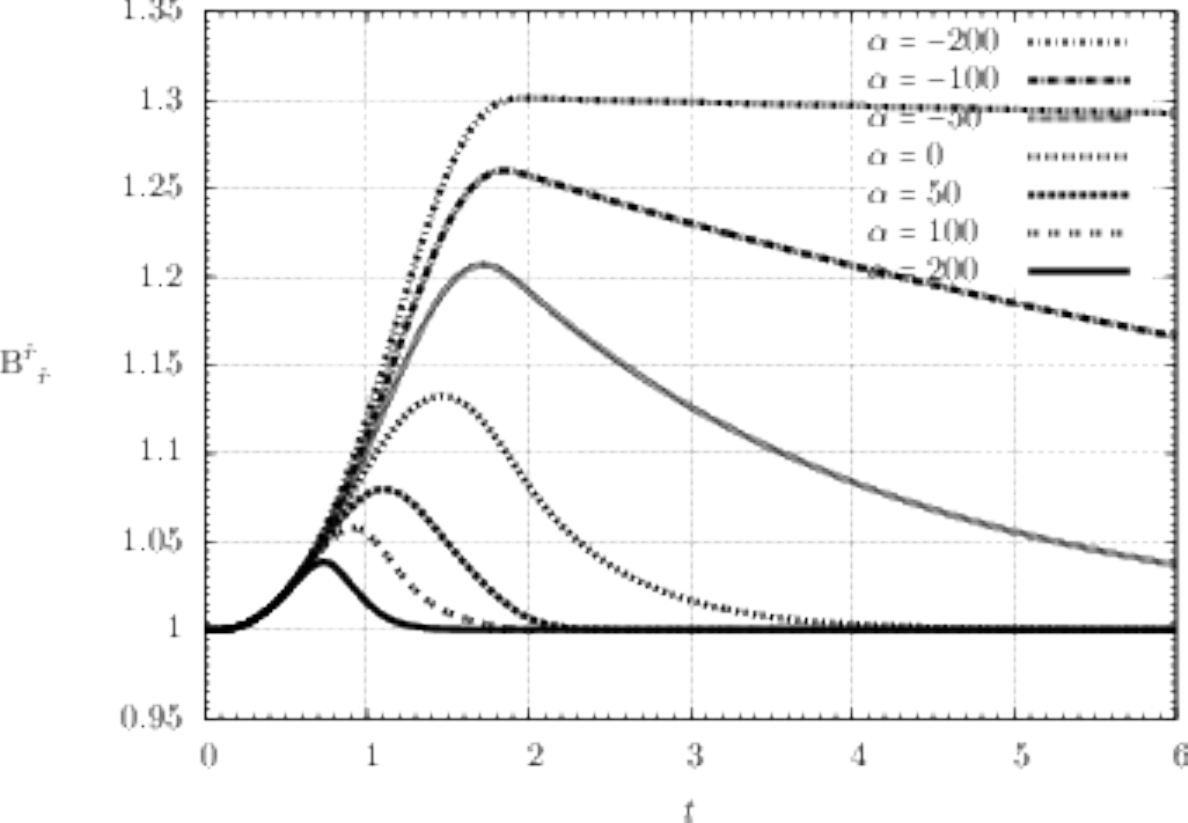}}
  \qquad
  \subfloat[\label{fig:response-b-lcgncBzz} Left Cacuhy-Green tensor $\lcgnc$, component $\Bzzhat$.]{\includegraphics[width=0.4\textwidth]{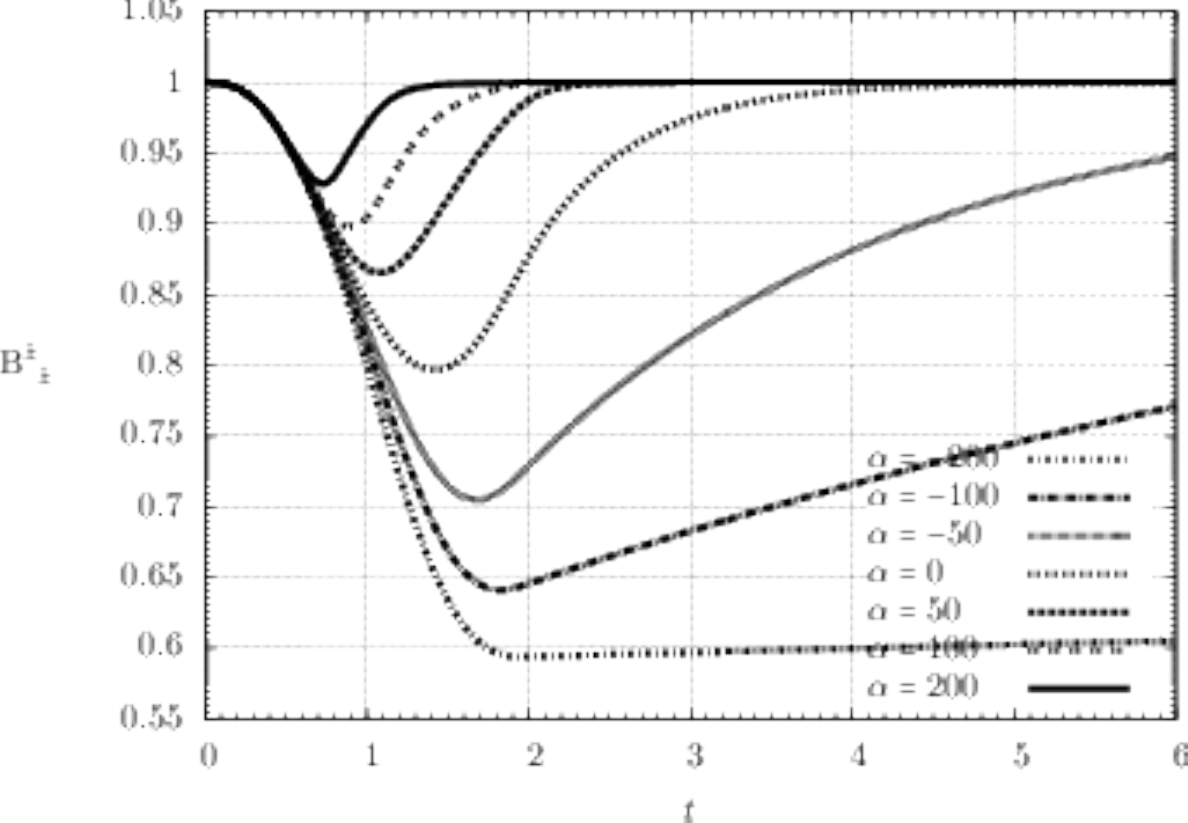}}
  \caption{Time evolution of $\temp$, $\Brrhat$, $\Bzzhat$ and $\entropy$ fields in the sample subject to biaxial extension induced by $h(t)$ shown in Figure~\ref{fig:sample-height}. Parameter values are $\temp_{\reference}=1$, $\mu_{\reference}=\sqrt{5}$, $\nu_1=\sqrt{2}$, $\nu = 100\sqrt{7}$, $\rho=1$ and $\cheatvol^{\mathrm{iNSE}}=1000 \sqrt{3}$.}
  \label{fig:response-b}
\end{figure}

In the other case, see Figure~\ref{fig:response-b}, the ``solvent'' viscosity $\nu$ is chosen to be relatively high. Consequently, the corresponding term on the right hand side of~\eqref{eq:114} dominates the other terms that depend on $\mu$. (See the caption of Figure~\ref{fig:response-b} for the values of the material coefficients.) This means that the ``elastic'' contribution should be negligible in the evolution equation for the temperature~\eqref{eq:114}. This is indeed the case, the temperature evolution is almost insensitive to the value of the material coefficient $\alpha$, see Figure~\ref{fig:response-b}. Moreover, the temperature behaves as one might expect, it is an increasing function of time. Similarly, the entropy evolution is also almost insensitive to the specific value of~$\alpha$, see Figure~\ref{fig:response-b-entropy}.

Further, the left Cauchy--Green tensor field $\lcgnc$ relaxes to the equilibrium value $\lcgnc = \identity$ as expected. However, the transitional behaviour again strongly depends on the value of the coefficient $\alpha$. Unlike in the heat equation~\eqref{eq:114}, the temperature variations of $\mu$ can not be neglected in the mechanical part of the system, that is in~\eqref{eq:112} and \eqref{eq:113}. 

\section{Numerical solution of governing equations}
\label{sec:numerical-solution}

\newcommand{\I}{{\bf I}}
\newcommand{\F}{{\bf F}}

Analysis of flows in complex domains and flows induced by complex boundary conditions requires one to develop numerical methods for the solution of the governing equations. Concerning the \emph{mechanical} part of the system of governing equations, the corresponding numerical methods have been subject to numerous studies, see for example \cite{crochet.mj.davies.ar.ea:numerical} and \cite{owens.rg.phillips.tn:computational} and references therein. The \emph{coupled thermomechanical} system has been studied far less frequently, see for example \cite{damanik.h.hron.j.ea:monolithic} and \cite{damanik.h:fem}, while the temperature evolution equation has been mainly considered in the \emph{naive} form~\eqref{eq:17} or~\eqref{eq:80}.

Below we document that the \emph{full} system of governing equations~\eqref{eq:maxwell-oldroyd-temperature-dependent-summary} for Maxwell/Oldroyd-B fluid with temperature dependent material coefficients can be also easily treated numerically in a relatively complex settings. (This is of course true only for low to moderate Weissenberg number and low Reynolds number. Out of this regime, more sophisticated numerical approaches are needed, see~\cite{fattal.r.kupferman.r:time-dependent} or a recent contribution by~\cite{sousa.rg.poole.rj.ea:lid-driven} and the references therein.) In particular we solve the governing equations in a setting that leads to a non-uniform temperature field generated by the dissipation. 

\begin{figure}[h]
  \centering
  \includegraphics[width=0.4\textwidth]{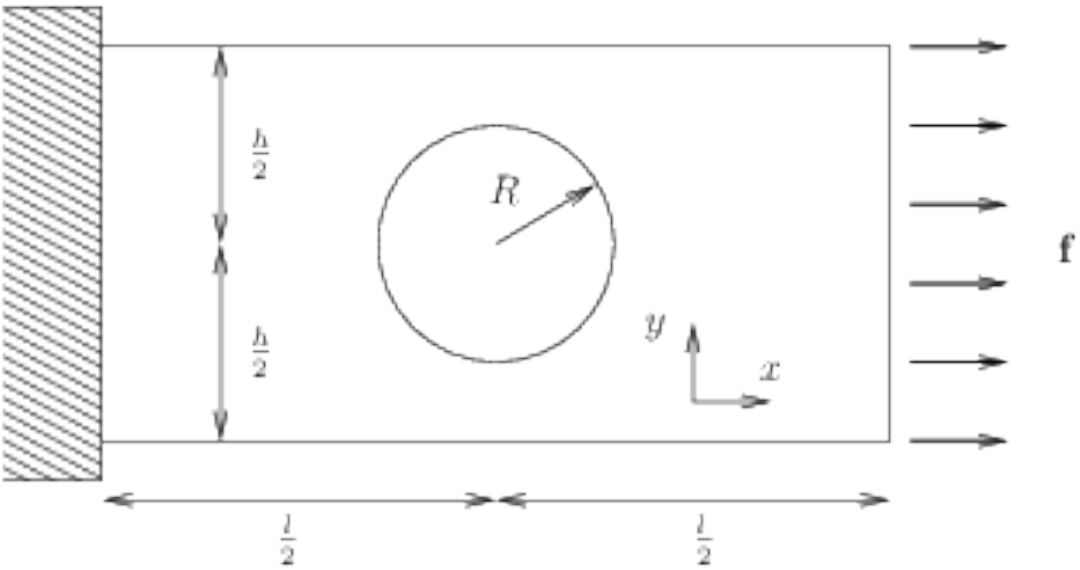}
  \caption{Problem geometry.}
  \label{fig:numerical-problem-geometry}
\end{figure}

We consider a two dimensional problem of a deformation of a rectangular plate with a circular hole, see Figure~\ref{fig:numerical-problem-geometry}. The plate is held fixed on the left boundary and an oscillating force $\vec{f}$ is acting on the right boundary. The top and bottom boundary and the boundary of the circular hole are traction free, and they are assumed to be thermally insulated. The left and right boundary are held at constant temperature $\temp_{\mathrm{left}}$ and $\temp_{\mathrm{right}}$ respectively. The acting force induces a deformation of the plate which successively leads to the heat generation in the plate. From the mathematical point of view, the additional complexity arises from the fact that one has to study the motion of the fluid in an \emph{a priori} unknown moving domain.   

In order to simulate such a flow the arbitrary Lagrangian--Eulerian (ALE) method is employed, see for example~\cite{donea.j.huerta.a.ea:arbitrary} and~\cite{scovazzi.g.hughes.t:lecture}. Our implementation is based on the work by~\cite{hron.j.rajagopal.kr.ea:flow}, who have considered the finite element method for numerical solution of the \emph{mechanical} part of the system of governing equations in a moving domain~$\Omega_\chi$. The weak formulation of the additional \emph{temperature} evolution equation in ALE formulation reads, in the notation adopted in~\cite{hron.j.rajagopal.kr.ea:flow}, as follows
\begin{multline}
\int_{\Omega_\chi}\hat{J} \left[
      \rho \sheatcapvol^{\mathrm{iNSE}} 
      -
      \left[
        \frac{\temp}{2}
        \ddd{\mu}{\temp}
        \left(
          \Tr \lcgnc
          -
          3
          -
          \ln \det \lcgnc
        \right)
      \right]
    \right] \left(\pd{\theta}{t}+\left(\hat{\F}^{-1}\left(\vec{v}-\pd{\hat{\vec{u}}}{t}\right)\right)\,\cdot\,\nabla_{\chi}\theta\right) q_\theta
    \,{\rm d}\chi 
    + 
    \int_{\Omega_\chi} \hat{J}\kappa(\theta)\hat{\F}^{-1}\hat{\F}^{-\rm T}\nabla_\chi\theta \cdot \nabla_\chi q_\theta\,{\rm d}\chi\\
    -
    \int_{\Omega_\chi} \hat{J}
    \left[
      \frac{\nu(\theta)}{2}\Big|(\nabla_\chi\vec{v})\hat{\F}^{-1}+\hat{\F}^{-\rm T}(\nabla_\chi\vec{v})^{\rm T}\Big|^2
      +
      \frac{\temp}{2}\dd{\mu}{\temp}
      \tensordot{\traceless{\left(\lcgnc\right)}}
      {((\nabla_\chi\vec{v})\hat{\F}^{-1}+\hat{\F}^{-\rm T}(\nabla_\chi\vec{v})^{\rm T})}
    \right]q_\theta\,{\rm d}\chi\\
    -
    \int_{\Omega_\chi} \hat{J}\left[
      \frac{\mu}{2 \nu_1}
      \left(\mu - \temp \dd{\mu}{\temp}\right)
      \left(
        \Tr \lcgnc + \Tr \left( \inverse{\lcgnc} \right) - 6
      \right)
    \right] q_\theta\,{\rm d}\chi=0
\end{multline}
where $q_\theta$ denotes the admissible test function. Here $\Omega_\chi$ is the preimage of the deforming domain in the ALE frame, $\hat{\vec{u}}$ is the corresponding ALE displacement, $\hat\F$ is the associated deformation gradient $\hat{\F}={\bf I}+\nabla_\chi\hat{\vec{u}}$, $\hat{J}=\det{\hat{\F}}$ and $\nabla_\chi$ denotes the gradient operator in the ALE frame. The weak formulation of the mechanical part of the system of governing equations is identical to that used in~\cite{hron.j.rajagopal.kr.ea:flow}. Note that the weak formulation in a \emph{fixed} Eulerian coordinates---which is suitable for numerical simulations of internal flows in fixed domains---is just a special case of ALE formulation. Indeed, it is sufficient to fix the ALE displacement $\hat{\vec{u}}=\vec{0}$, which yields $\hat{\F}=\I$ and $\hat{J}=1$, and the weak formulation in the fixed Eulerian coordinates follows immediately.

The weakly formulated governing equations are discretised in space using the finite element method and the time derivatives are approximated with the backward Euler method. The two-dimensional domain $\Omega_\chi$ is approximated by $\Omega_{\chi, h}$ with a polygonal boundary, and $\Omega_{\chi, h}$ is discretized by regular triangles. The velocity $\vec{v}$ is approximated by $P_2$ elements, other unknowns $\lcgnc$ and $\theta$ are also approximated by $P_2$ elements, while the pressure $p$ is approximated by linear elements $P_1$. The nonlinearities are treated by the Newton method with an exact tangent matrix computed using an automatic differentiation method. The resulting set of linear equations is solved by direct solver. In our implementation we use the software components bundled in the FEniCS project, see \cite{aln-s.m.blechta.j.ea:fenics}. 

Parameter values used in the computation are the flowing. The plate length is fixed to $l=20$, the height of the plate is $h=4$ and the radius of the circular hole is $R=1$. Temperature values at the ends of the plate are $\temp_{\mathrm{left}}=\temp_{\mathrm{ref}}$ and $\temp_{\mathrm{right}}=\temp_{\mathrm{ref}}$. Material coefficients $\mu$, $\nu$ and $\nu_1$ are given as $\mu=4 + 200 \exponential{-(\temp - \temp_{\mathrm{ref}})}$, $\nu_1=100$ and $\nu=10$, $\kappa = 0.01$, $\cheatvol=0.1$ and $ \rho=0.1$. The initial conditions at $t=0$ are chosen as $\left. \lcgnc \right|_{t=0}=\identity$, $\left. \vec{v} \right|_{t=0}= \vec{0}$ and $\left. \temp \right|_{t=0}= 0$. The surface force acting on the right boundary takes the form
\begin{equation}
  \label{eq:122}
  \vec{f} 
  =
  \begin{bmatrix}
    1.1 \sin t \\
    0
  \end{bmatrix}
  .
\end{equation}
This surface force is related to the \emph{computational} domain $\Omega_{\chi}$, that is we prescribe $\left(\hat{J}\hat{\bf T}\hat{\F}^{-\rm T} \vec{n}_{\chi}\right)=\vec{f}$ on the right boundary of the computational domain $\Omega_{\chi}$. This is a natural Neumann type boundary condition in the ALE reformulation of the governing equations, see~\cite{hron.j.rajagopal.kr.ea:flow} for details. 

\begin{figure}[h]
  \subfloat[\label{fig:numerical-problem-temperature-field-full-a}Time $t=200 \Delta t$, $t\approx 3.3\, T$.]{\includegraphics[width=0.8\textwidth]{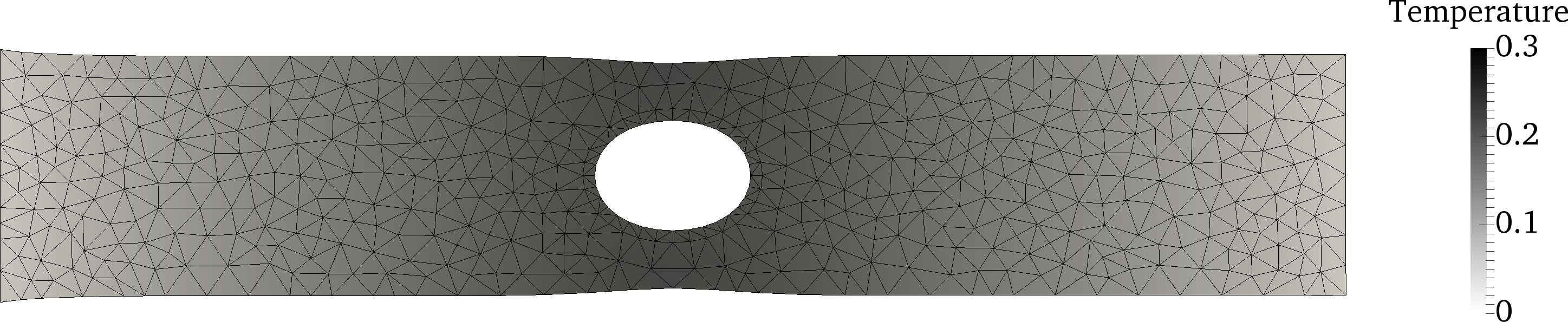}}
  \\
  \subfloat[\label{fig:numerical-problem-temperature-field-full-b}Time $t=492 \Delta t$, $t\approx 8.2\, T$.]{\includegraphics[width=0.8\textwidth]{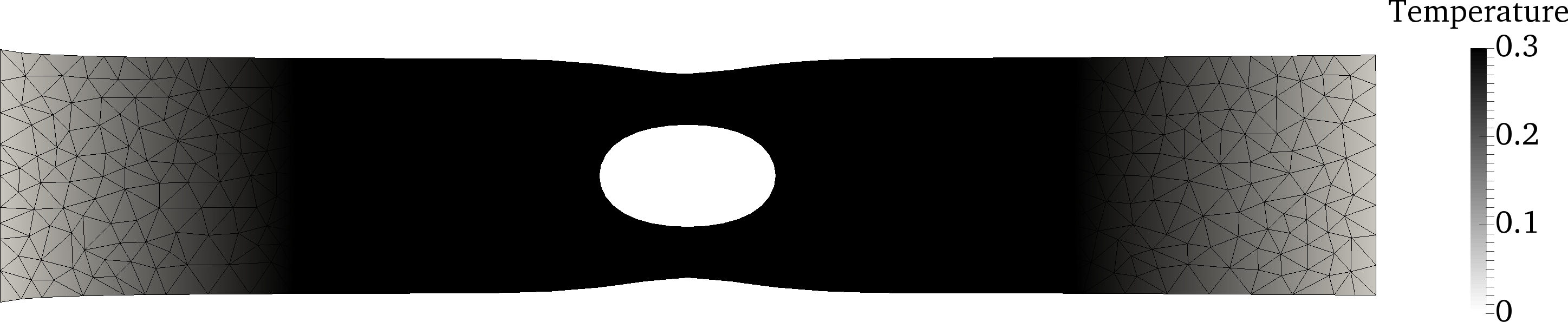}}
  \caption{Temperature field $\temp - \temp_{\mathrm{ref}}$ at various time instants -- full temperature evolution equation~\eqref{eq:77}. Displacement field is subject to forty times magnification.}
  \label{fig:numerical-problem-temperature-field-full}
\end{figure}

\begin{figure}[h]
  \centering
  \subfloat[\label{fig:numerical-problem-temperature-field-naive-a}Time $t=200 \Delta t$, $t\approx 3.3\, T$.]{\includegraphics[width=0.8\textwidth]{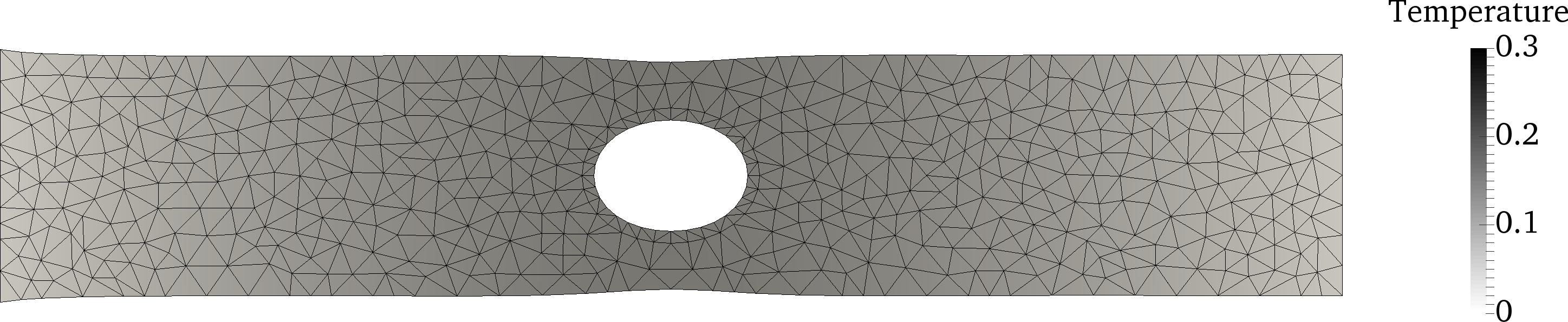}}
  \\
  \subfloat[\label{fig:numerical-problem-temperature-field-naive-b}Time $t=492 \Delta t$, $t\approx 8.2\, T$.]{\includegraphics[width=0.8\textwidth]{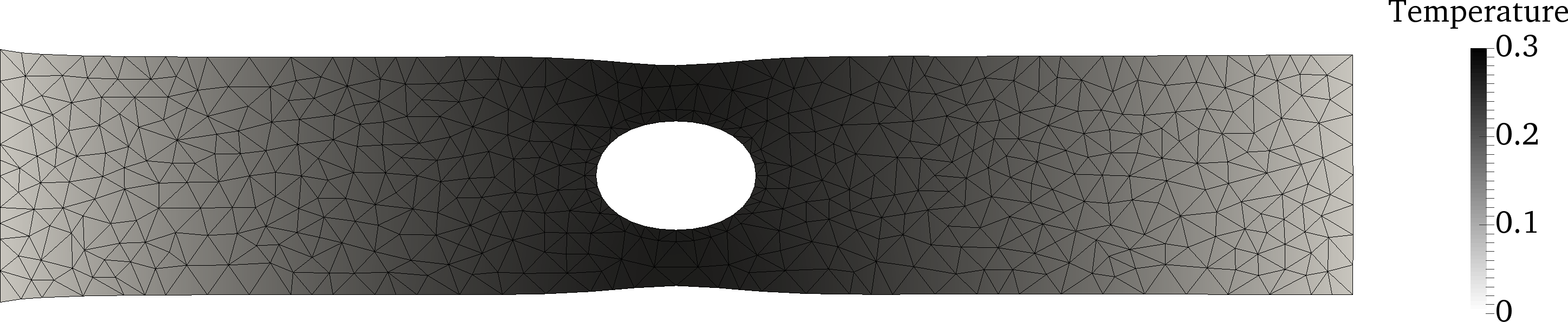}}
  \caption{Temperature field $\temp - \temp_{\mathrm{ref}}$ at various time instants -- naive temperature evolution equation~\eqref{eq:80}. Displacement field is subject to forty times magnification.}
  \label{fig:numerical-problem-temperature-field-naive}
\end{figure}

The computed inhomogeneous temperature field in the sample is shown in Figure~\ref{fig:numerical-problem-temperature-field-full} and Figure~\ref{fig:numerical-problem-temperature-field-naive}. Figure~\ref{fig:numerical-problem-temperature-field-full} shows the temperature field computed using the \emph{full} temperature evolution equation~\eqref{eq:77}, while Figure~\ref{fig:numerical-problem-temperature-field-naive} shows the temperature field computed using the \emph{naive} temperature evolution equation~\eqref{eq:80}.  In both cases we have used a mesh with $574$ elements, and the chosen time step is $\Delta t = 0.105$. 

As the sample oscillates with the period $T=2\pi$, an inhomogeneous temperature field develops in the material due to the dissipation, see Figure~\ref{fig:numerical-problem-temperature-field-full-a} and Figure~\ref{fig:numerical-problem-temperature-field-full-b} and Figure~\ref{fig:numerical-problem-temperature-field-naive-a} and Figure~\ref{fig:numerical-problem-temperature-field-naive-b}. The temperature increase predicted by the naive temperature evolution equation is however much smaller than that predicted by the full temperature evolution equation. Since $\mu$ decreases with temperature, the material loses stiffness with increasing temperature. Consequently, the sample is at the same time instant noticeably longer if one uses the full temperature evolution equation instead of the naive equation, compare Figure~\ref{fig:numerical-problem-temperature-field-full-b} and Figure~\ref{fig:numerical-problem-temperature-field-naive-b}. This again documents the thesis that the discrepancies in the temperature evolution equation can have significant impact on the predicted values of mechanical quantities.

\section{Conclusion}
\label{sec:conclusion}
We have investigated a class of models for fluids with temperature dependent viscoelastic type response. In particular, we have focused on Maxwell/Oldroyd-B type models with temperature dependent material coefficients. These classical models are in a sense generic models for non-Newtonian fluids, see~\cite{prusa.rajagopal.kr:on}, hence they constitute a natural starting-point for the detailed analysis of temperature induced effects in complex non-Newtonian fluids. Clearly, if one wants to successfully describe the motion of these fluids, one has to formulate not only governing equations for the mechanical quantities, but also for the temperature.

As we have argued, the correct temperature evolution equation is not straightforward to find. Since the temperature evolution equation is in fact an evolution equation for the thermal part of the internal energy, its correct formulation is tantamount to the correct description of the energy transfer mechanisms in the given fluid. However, the energy transfer mechanisms in a \emph{viscoelastic} fluid are quite complex, because one needs to describe the conversion between the kinetic energy, the stored energy in the elastic ``part'' of the fluid and the thermal energy. Consequently, the formulation of the correct temperature evolution equation is a delicate task, and a systematic procedure for the identification of the correct equation is necessary. 

Using a generalisation of a phenomenological approach by~\cite{rajagopal.kr.srinivasa.ar:thermodynamic} and \cite{malek.j.rajagopal.kr.ea:on}, we have found the appropriate temperature evolution equation for viscoelastic fluids described by Maxwell/Oldroyd-B models with temperature dependent material coefficients. Further, having the temperature evolution equation, we have formulated the complete system of governing equation for the mechanical and thermal variables. Note that besides the specific heat at constant volume $\cheatvol^{\mathrm{iNSE}}$ and the thermal conductivity $\kappa$, the final full system of governing equations does not include any additional material coefficients than those already present in the mechanical part of the system. In this sense, the derived full system of equations contains only material coefficients that are routinely measured, hence it can be directly used in practice. 

The derived correct evolution equation for the temperature reads
\begin{multline}
  \label{eq:120}
    \left[
      \rho \sheatcapvol^{\mathrm{iNSE}}
      -
      \left[
        \frac{\temp}{2}
        \ddd{\mu}{\temp}(\temp)
        \left(
          \Tr \lcgnc
          -
          3
          -
          \ln \det \lcgnc
        \right)
      \right]
    \right]
    \dd{\temp}{t}
    \\
    =
    2 \nu(\temp) \tensordot{\traceless{\gradsym}}{\traceless{\gradsym}}
    +
    \divergence \left(\kappa(\temp) \nabla \temp \right)
    +
    \temp\dd{\mu}{\temp}
    \tensordot{
      \traceless{\left(\lcgnc\right)}
    }
    {\traceless{\gradsym}}
    +
    \frac{\mu(\temp)}{2 \nu_1(\temp)}
    \left(
      \mu(\temp) - \temp \dd{\mu}{\temp}(\temp)
    \right)
    \left(
      \Tr \lcgnc + \Tr \left( \inverse{\lcgnc} \right) - 6
    \right),
\end{multline}
and it is apparently more complex than the naive heat equation 
\begin{equation}
  \label{eq:119}
  \rho \sheatcapvol^{\mathrm{iNSE}} \dd{\temp}{t}
  =
  \divergence \left(\kappa(\temp) \nabla \temp \right) + 2 \nu(\temp) \tensordot{\traceless{\gradsym}}{\traceless{\gradsym}},
\end{equation}
that is frequently used in practice. However, it is easy to see that the correct temperature equation~\eqref{eq:120} reduces well, under certain circumstances, to the naive evolution equation~\eqref{eq:119}. In this sense, equation~\eqref{eq:120} is an appropriate formulation of the naive equation~\eqref{eq:119}. On the other hand, there are also flow regimes in which the additional terms in the correct equation are important and can not be ignored. The impact of the additional terms on the flow dynamics has been documented by solution of simple boundary value problems. 

The approach used for the derivation of the full system of governing equations can be easily modified and generalised. In particular, a different \emph{ansatz} for the free energy or more complicated kinematics of the evolving natural configuration would lead to more involved viscoelastic rate type models. Further, the existing \emph{isothermal compressible} viscoelastic rate type models, see for example~\cite{bollada.pc.phillips.tn:on}, can be converted to the full thermomechanical setting as well. This in principle allows one to develop models that have the potential to describe quite complex temperature dependent rheological behaviour in a thermodynamically consistent manner.

\bibliographystyle{chicago}
\bibliography{vit-prusa}
\end{document}